\documentclass[aps,pra,twocolumn,floatfix,longbibliography]{revtex4-1}
\usepackage{amsmath}
\usepackage{amssymb}
\usepackage{graphicx}
\usepackage{bm}
\usepackage{xcolor}
\usepackage{ulem}
\usepackage{mathtools}
\graphicspath{{Pictures/}}
\usepackage{sidecap}
\usepackage{glossaries}
\makeglossaries

\newcommand{\qv}{{\mathbf v}}

\begin{document}

\title{Molecular Dynamics Study of the Sonic Horizon of Microscopic Laval Nozzles}

\author{Helmut Ortmayer$^{1,2}$, Robert E. Zillich$^1$}

\affiliation{$^1$Institute for Theoretical Physics, Johannes Kepler University,
  Altenbergerstrasse 69, 4040 Linz, Austria}
\affiliation{$^2$Primetals Technologies Austria GmbH, Turmstrasse 44, A-4031 Linz, Austria}

\begin{abstract}
  A Laval nozzle can accelerate expanding gas above supersonic velocities, while cooling the gas
  in the process. This work investigates this process for microscopic Laval nozzles
  by means of non-equilibrium molecular dynamics simulations of statioary flow,
  using grand canonical Monte-Carlo particle reservoirs. We study the expansion of a simple fluid,
  a mono-atomic gas interacting via a Lennard-Jones potential, through an idealized nozzle
  with atomically smooth walls.
  We obtain the thermodynamic state variables pressure, density, and temperature,
  but also the Knudsen number, speed of sound, velocity, and the corresponing Mach number
  of the expanding gas for nozzles of
  different sizes. We find that the temperature is well-defined in the sense that the
  each velocity components of the particles obey the Maxwell-Boltzmann distribution,
  but it is anisotropic, especially for small nozzles.
  The velocity auto-correlation function reveals a tendency towards condensation of the cooled
  supersonic gas, although the nozzles are too small for the formation
  of clusters. Overall we find that microscopic nozzles act qualitatively like macroscopic
  nozzles in that the particles are accelerated to supersonic speeds while their
  thermal motion relative to the stationary flow is cooled.
  We find that, like macroscopic Laval nozzles, microscopic nozzles also exhibit a sonic horizon,
  which is well-defined on a microscopic scale. The sonic horizon is positioned only slightly further
  downstream compared to isentropic expansion through macroscopic nozzles, where the sonic
  horizon is situated in the most narrow part.
  We analyze the sonic horizon by studying spacetime density correlations, i.e.\ how thermal
  fluctuations at two positions of the gas density are correlated in time and find that
  after the sonic horizon there are indeed no upstream correlations on a microscopic scale.
\end{abstract}

\maketitle

\printglossaries

\newglossaryentry{dsmc}{name=DSMC, description={\textbf{D}irect \textbf{S}imulation \textbf{M}onte \textbf{C}arlo: a probabilistic method for solving the Boltzmann equation for rarefied gas flows}}
\newglossaryentry{md}{name=MD, description={\textbf{M}olecular \textbf{D}ynamics: Simulation method for N-body atomic simulations according to Newton's equation of motion}}
\newglossaryentry{lammps}{name=LAMMPS, description={\textbf{L}arge-scale \textbf{A}tomic/\textbf{M}olecular \textbf{M}assively \textbf{P}arallel \textbf{S}imulator: an open source classical gls{md} molecular dynamics code \url{http://lammps.sandia.gov/} \cite{LAMMPSwwwpage,frenkel2001understanding}}}
\newglossaryentry{lj}{name=LJ, description={\textbf{L}ennard-\textbf{J}ones potential: a simple approximation of the potential of neutral atoms proposed by John Lennard-Jones}}
\newglossaryentry{nemd}{name=NEMD, description={\textbf{N}on-\textbf{E}quilbrium \textbf{M}olecular \textbf{D}ynamics: Same method as \gls{md} but applied on systems which are not in a equilibrated state}}
\newglossaryentry{gcmc}{name=GCMC, description={\textbf{G}rand \textbf{C}anonical \textbf{M}onte \textbf{C}arlo exchange of particles: Combined Monte Carlo and molecular dynamics method to simulate a grand canonical ensemble, implemented in \gls{lammps} \cite{frenkel2001understanding,LAMMPSwwwpage,frenkel2001understanding}}}
\newglossaryentry{vacf}{name=VACF, description={\textbf{V}elocity \textbf{A}uto \textbf{C}orrelation \textbf{F}unction: A self correlation function of velocities as function of the time.}}
\newglossaryentry{mc}{name=MC, description={\textbf{M}onte \textbf{C}arlo is a simulation method relying on random sampling and specifies a brought class of algorithms.}}

\section{Introduction}

\begin{figure*}[htbp]
  \raisebox{0.15cm}{
  \begin{minipage}{0.43\columnwidth}
    \includegraphics[width=1\columnwidth]{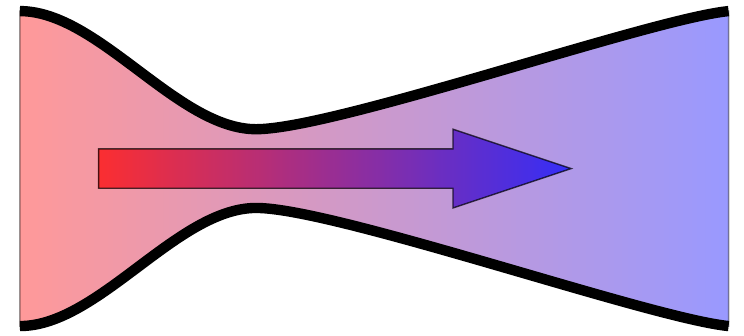}
  \end{minipage}}
  \begin{minipage}{0.66\textwidth}
    \includegraphics[width=1\textwidth]{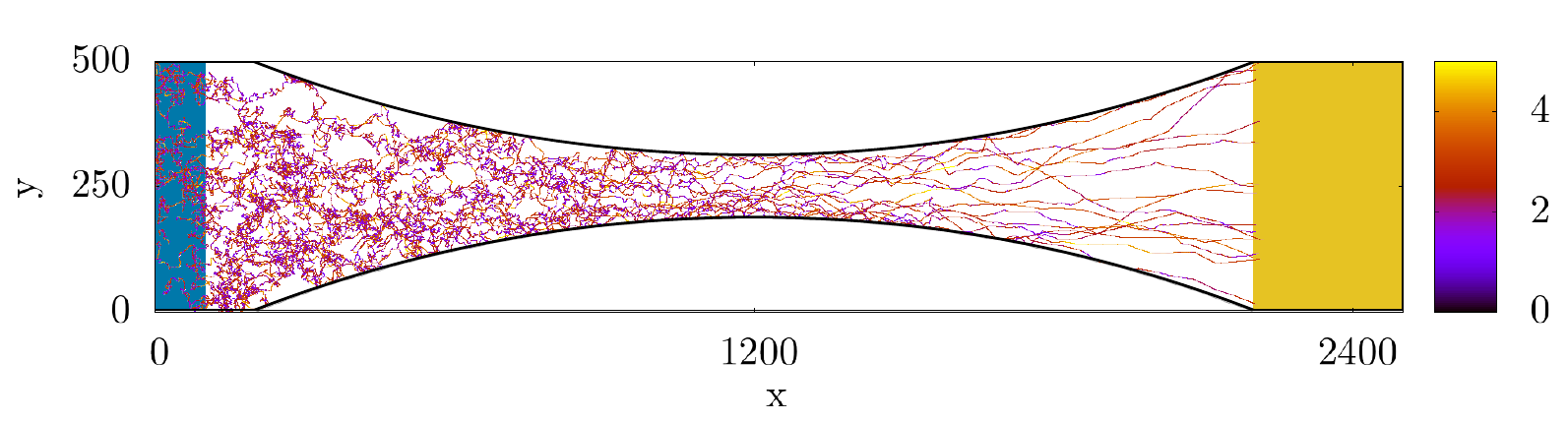}
  \end{minipage}
  \caption{Left: Cross section of a Laval nozzle with a convergent and divergent nozzle part.
    Indicated by the arrow and color is the flow direction and temperature decrease
    of the expanding gas.\\
    Right: Molecular dynamics trajectories of 30 randomly chosen particles starting in the shaded area to
    the left. The average total particle number in the nozzle for this simulation
    is much larger, approx.~790000.
    The velocity of these particles is indicated by color. While the subsonic motion in the convergent
    part is dominated by random thermal motion, the supersonic motion of the particles
    in the divergent part is }
  \label{picRandowWalk}
\end{figure*}

The Laval nozzle converts thermal kinetic energy into translational
kinetic energy and was invented by Gustaf de Laval in 1888 for actuating steam
turbines with steam accelerated by expansion. The goal was to achieve the
highest possible velocity of an expanding gas, made possible with the convergent-divergent
nozzle shape. The left panel of Fig.~\ref{picRandowWalk} schematically shows the
cross section of such a nozzle. When the gas reaches the most narrow part, the nozzle
throat, the flow can become supersonic. The surface where this
happens is called sonic horizon (or
acoustic horizon)~\cite{unruh1981experimental,visser1998acoustic} because no
information carried by sound waves can travel upstream through the sonic horizon.

The expansion of gas in a Laval nozzle has interesting thermodynamic
properties. While the gas acceleration of macroscopic Laval nozzles
is exploited for propulsion purposes in
rocket engines, the temperature drop during expansion through
a nozzle with a diameter in the tenth of $\mu$m range
is exploited in supersonic jet spectroscopy to freeze out translational, rotational
and vibrational degrees of freedom of molecules, leading to spectra that are not
complicated by too many thermally populated excited states
\cite{kantrowitz1951high,fitch1980fluorescence,smalley1977molecular,gough1977infrared}.
The studied molecules can be kept in a supercooled gas phase, far below the condensation
temperature, with a high density compared to a conventionally cooled equilibrium vapor.
Under appropriate conditions, weakly bound van der Waals cluster can be formed
\cite{skinner1980spectroscopy,johnston1984supersonic}. The molecules of interest are
typically co-expanded with a noble gas.
In case of $^4\mathrm{He}$ as carrier the cooling effect is also greatly enhanced by
the unique quantum effects of $^4\mathrm{He}$ at low temperatures. Especially the
helium-droplet beam technique takes additional advantage from the superfluidity
of $^4\mathrm{He}$\cite{toennies2004superfluid,skinner1980spectroscopy,johnston1984supersonic}.
The typical orifice used for molecular beams has only a convergent part and
the divergent nozzle part is realized by the ambient pressure in the expansion chamber.
During expansion the surrounding gas in the chamber provides a pressure boundary to the jet
and the jet temperature itself keeps decreasing after exiting the orifice.
\cite{sanna2005introduction}. 

Macroscopic Laval nozzles are well understood and can be approximately described by simple
thermodynamic considerations,
under assumptions that are reasonable for macroscopic nozzles: isentropic flow
without dissipation (inviscid gas and smooth slip boundaries); the flow velocity $v$
depends only on the position $x$ along the axis of the nozzle; the nozzles
cross section varies only gradually with $x$; the flow is stationary; and continuum fluid dynamics
is valid, i.e.\ each fluid element is in local thermodynamic equilibrium. Then
the relative velocity change with $x$ and the relative change of the cross section
area $A$ follow the simple relation \cite{sanna2005introduction}
\begin{equation}\label{equDivChangeVel}
  \frac{\mathrm{d}v}{v} =
  - \,\frac{1}{1-\left({\displaystyle v\over \displaystyle c}\right)^2}\,\frac{\mathrm{d}A}{A}
\end{equation}
where $c$ is the speed of sound, which can be expressed in terms of the isentropic or
isothermic derivative of the pressure with respect to the density,
\begin{equation}\label{equDefSpeedOfSound}
  c=\sqrt{\left(\frac{\partial p}{\partial \rho}\right)_S}
  = \sqrt{\frac{c_\mathrm{p}}{c_\mathrm{v}}\left(\frac{\partial p}{\partial \rho}\right)_T}
\end{equation}
where $c_\mathrm{p}$ and $c_\mathrm{v}$ is the heat capacity at constant pressure and
volume, respectively.
The ratio $M=v/c$ is called Mach number, and $M=1$ defines the sonic horizon.
The usual situation is a gas in a reservoir or
a combustion chamber producing gas to the left in our figures of the nozzle. Hence the
flow velocity is small when it enters the nozzle, in particular it is subsonic, $M<1$.
Eq.~(\ref{equDivChangeVel}) tells us that, with decreasing cross section $A$ (e.g. moving
downstream in the convergent part), the flow velocity $v$ must increase.
In the nozzle throat, i.e. where $A$ has a minimum and $dA=0$, $v$ either
stays below $M$, in which case $v$ must decelerate in the divergent part. Or the gas flow
attains $M=1$ in the nozzle throat, and then accelerates further in the divergent
part (if the pressure difference between inlet and outlet is large enough).
Hence for supersonic flow, $v$ increases with {\it increasing} $A$.
Note that Eq.~(\ref{equDivChangeVel}) implies that the transition to supersonic
flow can happen only where the cross section area has a minimum.

The goal of this work is to understand the physics of microscopic Laval nozzles
on the nanoscale of the atoms of the gas flowing through a constriction which is
only nanometers wide. We want to answer the following questions:
How do the transport properties of a Laval nozzle depend on its size, and does it even
have the typical characteristic of a convergent-divergent nozzle, i.e. converting
thermal energy into translational energy? If yes, how efficiently does a nanoscale
Laval nozzle cool the expanding gas? Do we obtain supersonic flow? 
Is there a well-defined sonic horizon, and if yes, where in the nozzle is it located?
Is there even local thermodynamic
equilibrium such that we can define a local speed of sound and
thus can speak of a sonic horizon and supersonic flow?
Since we are interested in the fundamental mechanism of a microscopic Laval nozzle
we study a rather idealized nozzle with atomically flat surfaces corresponding
to slip boundaries. This simplifies the problem since it eliminates the boundary
layer close to the nozzle walls. Boundary effects are of course essential in a real
microscopic nozzle, and they would be easy to model with rough walls,
but they would complicate the analysis and interpretation of our results.

A common method to study microscopic nozzles is the direct simulation Monte Carlo
(\gls{dsmc}) method \cite{boyd1992experimental,Horisawa200852,saadatiAST15,roohiPhysRep16}, which solves
the Boltzmann equation. However, we want to make as few approximations as possible,
apart from the idealization of a atomically smooth nozzle walls. Therefore we use
molecular dynamics (\gls{md}) simulations, which accounts for each atom or molecule
of the gas, and collisions are described by realistic intermolecular interactions.
Atomistic (\gls{md}) simulations have been shown to be useful for the understanding of fluid
dynamic phenomena~\cite{rapaportPRA87,moselerScience00,kadauPNAS04,horbachPRL06,yasudaPRX14,bordinJCP14,smithPhysFluids15,nowruziJMech18}.
The only underlying assumption of the \gls{md} method is that quantum physics plays no role
and classical mechanics is sufficient. This is usually a valid assumption, with the
exception of expansion of $^4\mathrm{He}$ under conditions
where the $^4\mathrm{He}$ gas cools to superfluid nanodroplets\cite{TVI98}.

Because of the non-equilibrium nature of this expansion process through a Laval nozzle 
we perform non-equilibrium \gls{md} (\gls{nemd}) simulations \cite{ciccotti2005non}.
The right panel of Fig.~\ref{picRandowWalk} shows the trajectories 30 randomly chosen
particles of a simulation in a convergent-divergent nozzle that contained
on average about 790000 particles. The speed of the particles is color-coded.
Fig.~\ref{picRandowWalk} gives an impression how a Laval nozzle converts thermal energy
(temperature) to ordered translation energy: close to the inlet the motion is predominantly
thermal; close to the outlet the velocities are higher and tend to point
in $x$-direction, but the temperature, i.e.\, the kinetic energy after subtracting
the flow velocity, is in fact much lower as our results will show.
Averaging over all particles and over time leads to the thermodynamic notion of a gas that
accelerates and cools as is expands through the nozzle.

With \gls{md} we can obtain, with microscopic resolution, both thermodynamic quantities like
temperature, pressure, or density, and microscopic quantities like the velocity
autocorrelation function \gls{vacf}, velocity distribution, or density fluctuation correlations:
we will investigate whether the expanding gas has a well-defined temperature, characterized by
an isotropic Maxwell-Boltzmann distribution of the thermal particle velocities.
The \gls{vacf} exhibits features related to the metastability of the
accelerated gas cooled below condensation temperature. We calculate spatio-temporal density
auto-correlations, i.e.\ correlations between fluctuations of the density at
different times and different locations, to study the propagation of information
upstream and downstream and pinpoint the location of the sonic horizon (if it exists).
In a macroscopic nozzle, upstream propagation of information carried by density fluctuations
is not possible in the supersonic region.
On the microscopic scale, e.g.\,on the scale of the mean free path of the atoms,
a unidirectional information flow is not so obvious. For instance, if we assume a
Maxwell-Boltzmann distribution of random particle velocities,
fast particles from the tail of the distribution could carry information upstream.

We remark that, in a seminal paper by W.~G.~Unruh et al.\ \cite{unruh1981experimental},
a mathematical analogue between the black hole evaporation by Hawking radiation and
the fluid mechanical description of a sonic horizon is found.
This analogue has brought significant attention to sonic horizons
\cite{garayZollerPRA01,steinhauer2015measuring,steinhauer2015observation,barcelo2011analogue,visser1998acoustic}, but in this work we will not study analog Hawking radiation.


\section{Molecular Dynamics Simulation of Expansion in Laval Nozzle}

The gas flow through the microscopic Laval nozzle is simulated with the molecular dynamics
(\gls{md}) method which solves Newton's equation of motion for all particles of the gas.
Unlike in continuum fluid dynamics, which solves the Navier-Stokes equation, 
\gls{md} contains thermal fluctuations of the pressure and density, also
in equilibrium. Furthermore, unlike the continuum description, \gls{md} does not assume
local thermodynamic equilibrium, which may not be fulfilled in a microscopic nozzle.

The price for an accurate atomistic description afforded by \gls{md} simulations
is a high computational cost compared to Navier-Stokes calculations or
\gls{dsmc} simulations.  In the present case, we simulate up to several hundred thousand
particles. Larger \gls{md} simulations are possible, but our focus is the
microscopic limit of a Laval nozzles on the nanometer scale.
A challenge for \gls{md} is to implement effective reservoirs to maintain a pressure differential
for a steady flow between inlet and outlet of the nozzle. An actual
reservoir large enough to maintain its thermodynamic state during the \gls{md} simulation would be prohibitively
computationally expensive. We approximate these reservoirs by defining small inlet and
outlet regions where we perform a hybrid \gls{md} and \gls{mc} Monte-Carlo simulation
(\gls{gcmc})~\cite{heffelfingerJCP94},
with grand canonical Monte-Carlo exchange of particles \cite{frenkel2001understanding}.
As the name implies, this method simulates a grand canonical ensemble for a given
chemical potential $\mu$, volume $V$ and temperature $T$ by inserting and removing particles.
The nozzle itself is simulated in the microcanonical ensemble, i.e.\ energy is conserved.
This ensemble represents a nozzle with perfect thermally insulating walls.

\begin{figure}[htbp]
	\begin{center}
	\includegraphics[width=1\columnwidth]{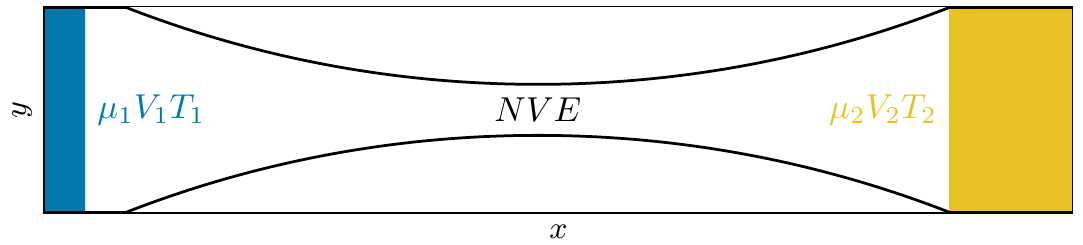}
	\caption{Geometry of a slit Laval nozzle with the convergent and divergent part
          in the $xy$-plane. The nozzle walls are two cylinders.
          In the $z$-direction out of the plane, the nozzle is translationally
          invariant, realized with periodic boundary conditions. Particle insertion
          is done by grand canonical Monte Carlo insertion and deletion
          \cite{heffelfingerJCP94,frenkel2001understanding} on the left side (blue) in the $\mu_1 V_1 T_1$
          ensemble.
          The nozzle region shown in white with the convergent end divergent boundaries is
          simulated in the microcanonical ensemble.
          Particle deletion is done on the right side (yellow) in a $\mu_2 V_2 T_2$
          ensemble.}
	\label{picSimulationEnsemble}
	\end{center}
\end{figure}

Fig.~\ref{picSimulationEnsemble} shows the geometry of the nozzle simulated
with the inlet and outlet colored in blue and yellow, respectively, with the
convergent-divergent nozzle in between. To keep the simulation simple and the computational
effort in check we simulate a slit Laval nozzle, translationally invariant in 
$z$-direction (perpendicular to the plane of the figure) and realized with
periodic boundaries in this direction.
Since our focus is a microscopic understanding of supersonic flow and the sonic
horizon, we simulate a nozzle with atomically smooth walls. Simulating rough walls
would have significantly complicated the analysis of the flow, because of the
nontrivial spatial dependence of the flow field in the direction perpendicular to
the general flow direction, requiring significantly longer simulations to
resolve all measured quantities in both $x$ and $y$ direction. In a smooth-walled
nozzle, we can restrict ourselves to studying only the $x$-dependence of the quantities
of interest.

The gas particles are atoms interacting via a pair-wise Lennard-Jones (\gls{lj}) potential. Thus
we simulate the expansion of a noble gas through the nozzle. Molecules with vibrational
and rotational degrees of freedom seeded into the noble gas would be an interesting
subject for further investigation, but this exceeds the scope of this work. The
(\gls{lj}) potential between a pair of particles with distance $r$ is given by
\begin{equation}\label{equLJ}
  V_{LJ}(r)=4\epsilon\left[ \left(\frac{\sigma}{r} \right)^{12}-\left(\frac{\sigma}{r}\right)^6 \right]
\end{equation}
The smooth walls are also modelled via a (\gls{lj}) potential with,
\begin{equation}\label{equLJWall}
  V_{LJ}(s)=4\epsilon\left[ \left(\frac{\sigma}{s} \right)^{12}-\left(\frac{\sigma}{s}\right)^6 \right]
\end{equation}
where $s$ is the normal distance between atom and wall.

We use the common reduced units for simulations of LJ particles if not otherwise stated,
see table~\ref{tabRedUToSi}.
Thus with the atom mass $m$, and the \gls{lj} parameters $\sigma$ and $\epsilon$
for a specific noble gas, the results can be converted from reduced units
to physical units.

\begin{table}[htbp!]
\caption{Conversion to dimensionless reduced units ($^*$) used in this work.}
\label{tabRedUToSi}
\begin{center}
\begin{tabular}{|ll|}
\hline 
Quantity & reduced units \\ 
\hline 
Distance & x$^*$=x/$\sigma$ \\ 
Time & t$^*$=$t\sqrt{\frac{\epsilon}{m^* \sigma^2}}$ \\ 
Energy & E$^*$=$E/\epsilon$ \\ 
Velocity & v$^*$=$\mathrm{v}\mathrm{t}^*/\sigma$ \\ 
Temperature & T$^*$=T $k_B / \epsilon$ \\ 
Pressure & P$^*$=P $\frac{\sigma^3}{\epsilon}$ \\ 
Density & $\rho^*$=$\rho \sigma^3$ \\ 
\hline 
\end{tabular} 
\end{center}
\end{table}

Atoms are inserted and deleted in the inlet (blue) and outlet (yellow) by running
the \gls{md} simulation in these regions as a hybrid (\gls{gcmc}) simulation~\cite{heffelfingerJCP94}.
The two grand canonical ensembles are characterized by their
chemical potential, the volume, and the temperature, $(\mu_1, V_1, T_1)$
and $(\mu_2, V_2, T_2)$, respectively. A proper choice of these thermodynamic
variables ensures that on average, an excess of particles are inserted in the inlet
and particle are eliminated in the outlet, such that a stationary gas flow is established
after equilibration. There are alternative insertion method, such as the
insertion-deletion method, where the mass flow is specified~\cite{barclayPRE16}.

The temperature and chemical potential of the inlet reservoir is set
to $T_1=2.0$ and $\mu_1=-32$, which would correspond to a density $\rho_1=0.86$ and ensures
that the pressure is not too high and the LJ particles remain in the gas phase.
The particle insertion region in the nozzle is not in equilibrium with the
grand canonical reservoir defining the $(\mu_1, V_1, T_1)$ ensemble, because the inlet volume
is not closed on the side facing the nozzle. The outflow must be compensated by additional insertions,
which makes the insertion rate higher than the elimination rate. Indeed we observed that
the average density in the insertion region is approximately half the density $\rho_1$.
Also the temperature in the inlet region is lower than the set value $T_1=2.0$.
The resulting pressure in the insertion region is $p\approx 0.06$ in our reduced units.
For Argon with $\epsilon=1.65\cdot 10^{-21}\,\mathrm{J}$ and
$\sigma=3.4$\,\AA\ \cite{griebel1997numerical} this translates to a temperature of $T=179$K
and a pressure $p\approx2.5\cdot 10^{6}\,\mathrm{Pa}$ in SI units. This is in the
pressure range for molecular beam spectroscopy experiments \cite{fitch1980fluorescence}.

The inlet conditions will converge to the specified reservoir variables if the number of
\gls{gcmc} moves is significantly larger than the number of \gls{md} moves, or if the
size of the inlet region is increased; both increases computational cost.
Alternatively, the inlet conditions may be matched to the desired pressure and temperature
by fine-tuning the reservoir variables and running many equilibration simulations,
which again requires a high computational effort. In this work
we refrain from perfectly controlling the thermodynamic state of the inlet although
it leads to effectively different inlet conditions in differently sized nozzles. 

In the convergent-divergent part of the nozzle, between the two grand canonical ensembles,
the atoms are propagated in the microcanonical ensemble (i.e. energy and particle number
are conserved), which is the most suitable ensemble for dynamic studies since the dynamics is not
biased by a thermostat. Since we want to simulate expansion into vacuum,
instead of choosing a very negative chemical potential,
we simply set the pressure in the outlet to zero, such that particles entering the outlet region
are deleted immediately.

For comparisons of different nozzle sizes, we scaled the slit nozzle in both
$x$ and $y$ directions, while keeping the simulation box length $z_{max}$ in the translationally invariant
$z$-direction, perpendicular to the figure plane in Fig.\,\ref{picSimulationEnsemble}, fixed.
In the $z$-direction, we apply periodic boundary conditions.
We compared different simulation box lengths $z_{max}$ in $z$-direction to quantify unwanted
finite size effects in $z$-direction. Ideally, we want to keep $z_{max}$ larger than the mean free path.
Especially for the dilute gas at the end of the divergent part, a sufficiently
large $z_{max}$ is required to avoid such effects. For most simulations,
we found $z_{max}=86.18\,\sigma$ or $z_{max}=43.09\,\sigma$ to be adequate, as shown below.

We initialize the \gls{nemd} simulations with
particles only in the inlet region. Equilibration is achieved when the total number
of particle in the simulation does not increase anymore but just fluctuates
about an average value. When this steady state is reached, we start measurements by
averaging velocities, pressure, density etc.

The equilibrium equation of state for LJ particles is well
known~\cite{johnson1993lennard,boda1996isochoric}. The equation of state is not needed for
the \gls{md} simulations, but it is helpful for the analysis of the results, particularly
for the calculation of the speed of sound and the Mach number.
Specifying the Mach number, temperature, or pressure rests on the assumption of
local thermodynamic equilibrium, and thus on the validity of a local equation of state.
In a microscopic nozzles where the state variables of the LJ gas changes on a
very small temporal and spatial scale local thermodynamic equilibrium may be violated.

All simulation were done with the open source \gls{md} software \gls{lammps}~\cite{plimpton1995fast,LAMMPS}.

\section{Thermodynamic properties}

In this section we present thermodynamic results of our molecular dynamic simulations of
the expansion through slit Laval nozzles: density, pressure,
temperature, and Mach number. We check whether a microscopic nozzle exhibits
the transition to supersonic flow and where the sonic horizon is located in nozzles
of various sizes, and we compare to ideal gas continuum dynamics.
The atomistic \gls{nemd} simulation also allows us to investigate if the gas
attains a local equilibrium everywhere in the nozzle, with a well-defined temperature.

\begin{figure}[htbp]
	\begin{center}
	\includegraphics[width=0.42\textwidth]{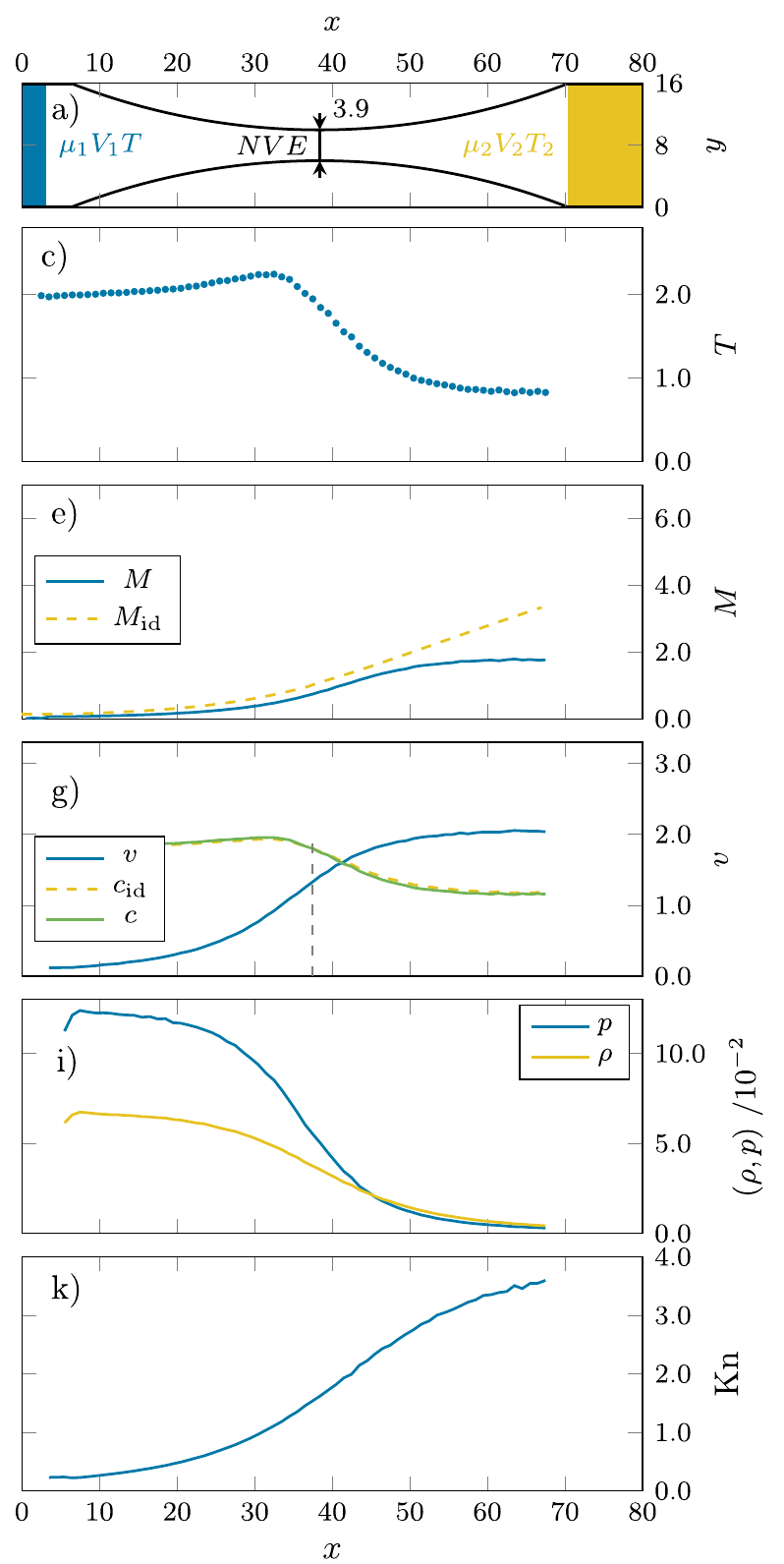}
	\caption{Thermodynamic quantities for a nozzle with a throat width of only
          $3.9\,\sigma$. The figure shows in panel a) an overview of the nozzle in the x-y-plane, in c)
          the temperature, in e) the Mach number $M(x)$ and the ideal gas approximation for
          the Mach number $M_\mathrm{id}$, in g) the ideal gas approximation of the speed of sound
          $c_\mathrm{id}$, the speed of sound $c$ obtained from the simulation and the averaged
          flow speed $v$, in i) the density $\rho$ and pressure $p$, and in k) the Knudsen number.
          All quantities are shown as a function of the $x$ position in the nozzle.}
	\label{picRasterAxis0p0312}
	\end{center}
\end{figure}

\subsection{Very small nozzle}

Fig.~\ref{picRasterAxis0p0312} shows results for a very small Laval nozzle,
with a throat width of only $3.9\,\sigma$, i.e. only a few atoms wide.
Panel a) shows the nozzle geometry.
The temperature is shown in panel c).
The kinetic temperature is the thermal motion of the atoms
after the flow velocity at ${\bf r}$, ${\bf v}({\bf r})$
is subtracted
\begin{equation}\label{eq:tempdef}
{3\over 2}k_\mathrm{B} T = \sum_i {m\over 2}\left({\bf v}_i - {\bf v}({\bf r_i})\right)^2
\end{equation}
Unlike in equilibrium, the temperature in a non-equilibrium situation such
as stationary flow varies spatially, $T=T({\bf r})$,
provided that local equilibrium is fulfilled. If there is no local equilibrium,
there is no well-defined temperature. Although the right hand side of
eq.(\ref{eq:tempdef}) can still be evaluated, the notion of a
``temperature'' is meaningless if the
thermal parts of the atom velocities do not follow a Maxwell-Boltzmann
distribution. Here we assume that eq.(\ref{eq:tempdef}) provides
a well-defined local temperature $T(x)$ at position $x$ along the flow
direction in our Laval nozzles. Further below we investigate whether this
assumption is justified.
The subtleties of the calculation of ${\bf v}({\bf r})$ and $T(x)$, and how
to subtract the flow velocity from the particle velocities can be found
in appendix~\ref{sec:appC} and ~\ref{sec:appD}, respectively.

Fig.~\ref{picRasterAxis0p0312} shows that
$T(x)$ indeed drops after the gas passes the nozzle throat, but there is a small
increase before it reaches the throat. We attribute this to the wall potential:
the constriction is dominated by the attractive well of the LJ potential
(\ref{equLJWall}). The associated drop in potential energy is accompanied
by an increase of the temperature, i.e. kinetic energy.

\begin{figure*}[htbp]
	\begin{center}
	\includegraphics[width=0.84\textwidth]{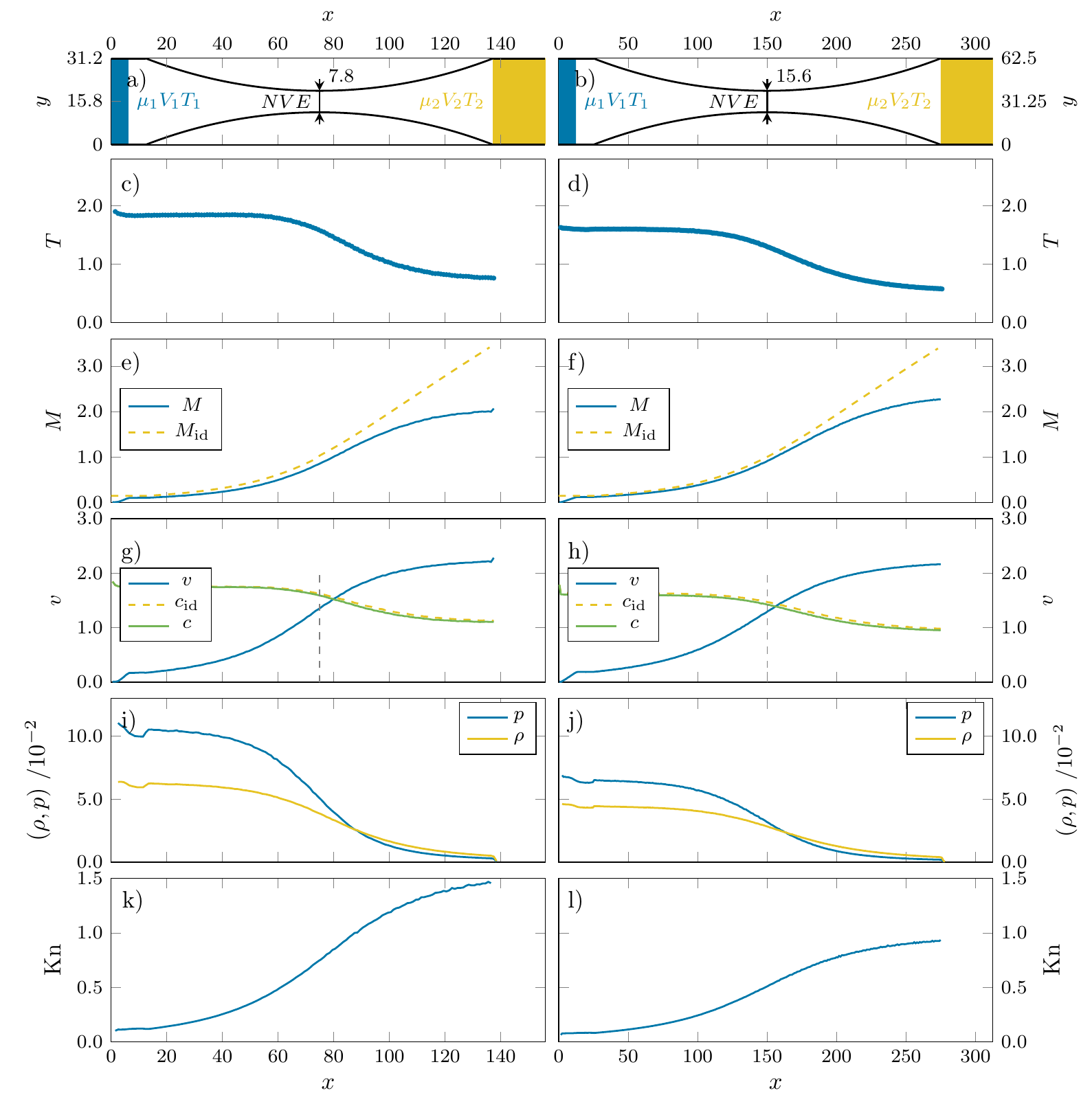}
	\caption{Same as Fig.\ref{picRasterAxis0p0312} for a throat width of
	$7.8\,\sigma$ (left column) and $15.6\,\sigma$ (right column).
	}
	\label{picRasterAxis0p1248}
	\end{center}
\end{figure*}

Panel g) shows the flow speed $v(x)=|{\bf v}(x)|$.
$v(x)$ increases monotonously over the whole length of the nozzle.
For comparions, we also show the speed of sound of the LJ gas $c(x)$ and of the ideal gas
$c_\mathrm{id}(x)$, which are very similar, even in the convergent part
where the density is higher.
For a monatomic ideal gas, the speed of sound (\ref{equDefSpeedOfSound}) becomes
\begin{equation}\label{eq:cid}
  c_\mathrm{id}(x)=\sqrt{{5\over 3}k_\mathrm{B} T(x) /m}.
\end{equation}
The speed of sound $c(x)$ of the LJ fluid is calculated from its equation of state
given in Ref.~\cite{johnson1993lennard} and the specific residual heat capacities
\cite{boda1996isochoric}, using the expression with the isothermal derivative
in Eq.~(\ref{equDefSpeedOfSound}) and the values of $\rho(x)$ and $T(x)$ measured in
the MD nozzle simulations. $\rho(x)$ is shown in panel i), together with the pressure.
The heat capacities $c_\mathrm{p}$ and $c_\mathrm{v}$ appearing in eq.(\ref{equDefSpeedOfSound})
are also obtained from the equation of state of the LJ fluid.
Note that applying the equation of state at position $x$ in the nozzle again assumes local
equilibrium, which is not necessarily true.

Panel e) shows the Mach number $M(x)$ obtained from the simulation and
the Mach number $M_\mathrm{id}(x)$ for an ideal gas continuum. For the ideal gas,
we can derive from eq.~(\ref{equDivChangeVel}) a relation between
the cross section areas $A(x)$ and Mach numbers $M_\mathrm{id}(x)$ at two
different positions $x_1$ and $x_2$ in the nozzle \cite{sanna2005introduction}
\begin{align}
  \frac{A(x_1)}{A_\mathrm(x_2)} &=
  \frac{M_\mathrm{id}(x_2)}{M_\mathrm{id}(x_1)}
  \left(\frac{1+\frac{\gamma -1}{2}M_\mathrm{id}^2(x_1)}{1+\frac{\gamma -1}{2}M_\mathrm{id}^2(x_2)}\right)^{\frac{\gamma+1}{2\left(\gamma-1 \right)}}
  \label{eq:TheoMachNumber}
\end{align}
$M_\mathrm{id}(x)$ can now be obtained by setting $x_1=x$ and $x_2=x_c$, the position of the
sonic horizon, where $M_\mathrm{id}(x_c)=1$ by definition.
Panel e) shows that the Mach number $M(x)$ obtained from the simulation
stays below the ideal gas approximation $M_\mathrm{id}(x)$,
with the difference growing in the divergent part of the nozzle. At the end of the nozzle
$M$ is approximately half the value of the ideal gas continuum approximation $M_\mathrm{id}$.
In particular, the sonic horizon predicted by the MD simulation
is located {\it after} the throat of the nozzle, not at the point of smallest cross section
predicted by the continuum description of isentropic flow, see eq.~(\ref{equDivChangeVel}).

\begin{figure*}[htbp]
	\begin{center}
	\includegraphics[width=0.84\textwidth]{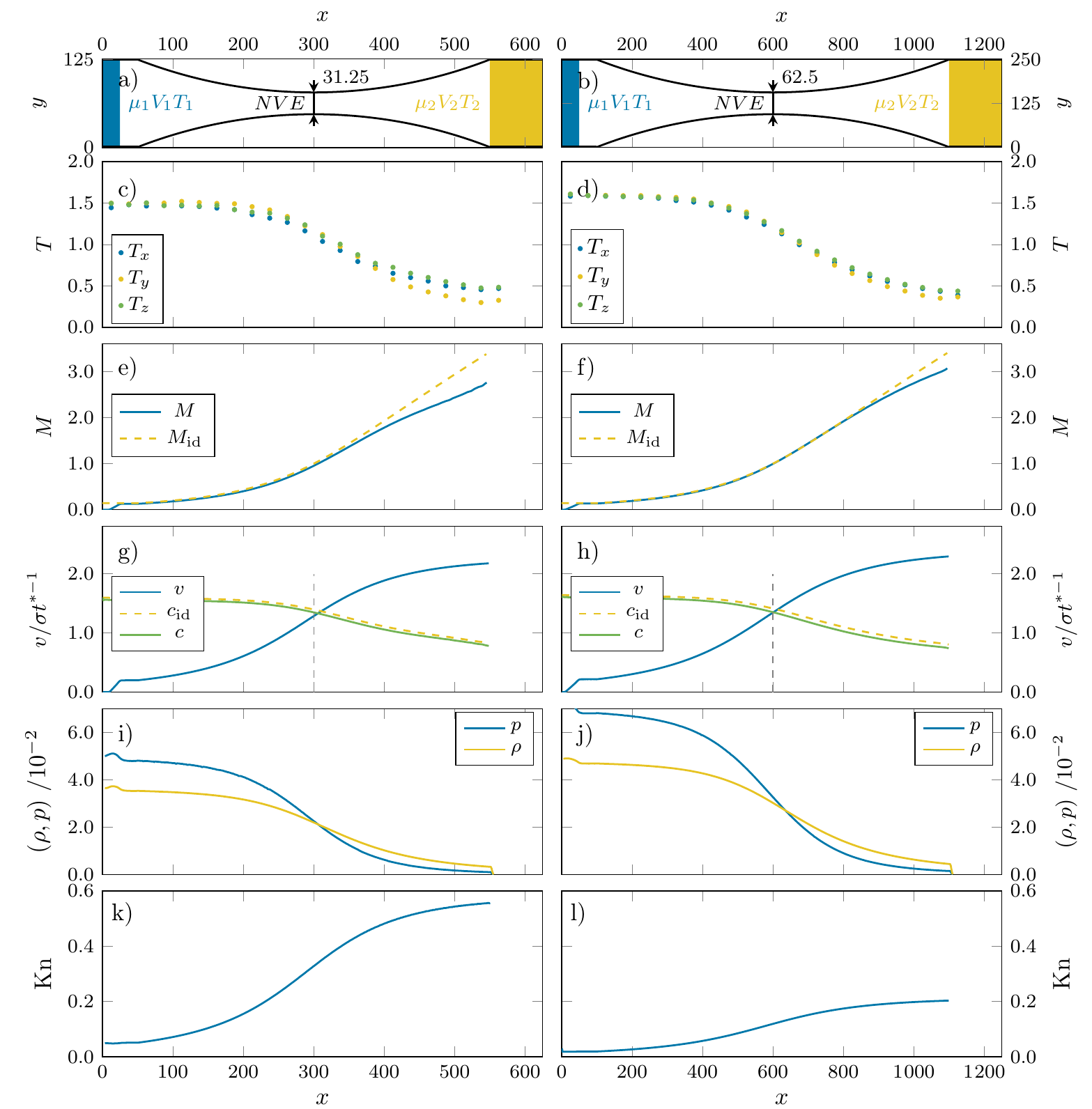}
	\caption{Same as Fig.\ref{picRasterAxis0p0312} for a throat width of
	$31.25\,\sigma$ (left column) and $62.5\,\sigma$ (right column). 
	The temperature is split into its contribution from motion in 
	$x$, $y$ and $z$ direction. }
	\label{picRasterAxis0p5}
	\end{center}
\end{figure*}

The Knudsen number is a characteristic quantity for flow in confined geometries.
It is the mean free path length $\lambda$ divided by a characteristic length $d$
of confinement
\begin{equation}\label{eq:KnudsenNumber}
	\mathrm{Kn}(x)={\lambda(x) \over d(x)}
\end{equation}
In our slit Laval nozzle $d(x)$ is the width at position $x$. We estimate the mean
free path $\lambda(x)$ using a hard sphere approximation
\cite{chapman1970mathematical}
\begin{equation}\label{eq:MeanFreePath}
  \lambda(x)=\left(\sqrt{2} \rho(x) \pi \right)^{-1}
\end{equation}
under the assumption of a Maxwell-Boltzmann distribution of the velocities
which which check to be fulfilled in the nozzle, see
section~\ref{sec:VelocityDistribution} and Fig.\,\ref{picVelDistribution}.
For $\mathrm{Kn}\ll$1 the mean free path is much smaller
than the nozzle width and a continuum description of the flow is appropriate.
For $\mathrm{Kn}\approx 1$ or $\mathrm{Kn} \gg 1$ a continuum description is
is not possible and the transport becomes partly ballistic.
For the smallest nozzle results, the Knudsen number $\mathrm{Kn}(x)$ shown
in panel k) in Fig.~\ref{picRasterAxis0p0312}, is significantly larger than unity
in the supersonic regime.

\subsection{Small nozzles}

Fig.~\ref{picRasterAxis0p1248} shows results for two nozzles twice and four times as
large as the smallest nozzle presented in Fig.\ref{picRasterAxis0p0312}, with
throat widths $7.8\,\sigma$ and $15.6\,\sigma$, respectively. The small
temperature increase seen for the smallest nozzle is not present anymore. $T$ is almost constant
in the convergent part and then decreases monotonously.
Note that for each nozzle, the flow starts from
slightly different thermodynamics conditions in the inlet region, for reasons
explained above. As the nozzle size increases, the Mach number $M$ reaches a higher
value for the larger nozzle despite the slightly lower $T$ in the inlet, and it follows
the ideal gas approximation $M_\mathrm{id}$ more closely. The sonic horizon moves closer
to the minimum of the cross section. Of course the Knudsen number $\mathrm{Kn}(x)$
is smaller for larger nozzles.
Due to the wider nozzle throat, the pressure is significantly lower in the convergent part.

For Fig.~\ref{picRasterAxis0p5}, we increase the nozzle size again twofold and fourfold.
We find the same trends as in Fig.~\ref{picRasterAxis0p1248}. For the nozzle
with throat width $62.5\sigma$, the Mach number $M$
is close to the ideal gas approximation $M_\mathrm{id}$. $M$ falls below
$M_\mathrm{id}$ only towards the end of the nozzle, where the collision rate
presumably becomes too low for efficient cooling. 
The sonic horizon is essentially in the center, indicated by the vertical dashed line.

For these two largest nozzles, we examined whether local equilibrium is fulfilled.
The direction-dependent temperature, see appendix~\ref{sec:appD}, is shown in
panel c) and d) of Fig.~\ref{picRasterAxis0p5}. The temperature is not quite isotropic,
i.e.\,there is insufficient local equilibration between the motion in $x$-, $y$, and
$z$-direction. The three respective temperatures differ. In the convergent part the
temperature in the $y$-direction, $T_y$, is highest, while
in the divergent part $T_y$ is lower than $T_x$ and $T_z$. $T_z$ is only influenced
by collisions between particles because there is no wall in $z$-direction. Comparing
the two nozzles presented in Fig.~\ref{picRasterAxis0p5}, we observe the expected
trend that the temperature anisotropy decreases with increasing nozzle size.
At the end of the nozzles in Fig.~\ref{picRasterAxis0p5} the temperature anisotropy
grows because the collision rate between particles drops as the density drops.
Whether the random
particle velocities are Maxwell-Boltzmann distributed will be studied in section
\ref{sec:microscopic} about microscopic properties.

In table~\ref{tab:AbsolutePosSonicHorizon} we compare the difference $\Delta x_c=x_c-x_c^0$ between
the calculated position $x_c$ of the sonic
horizon and the position $x_c^0$ of minimal cross section area
predicted by isentropic flow in the continuum description. In all cases the sonic horizon
is ``delayed'' and shifted downstream, $\Delta x_c>0$. With growing nozzle size
characterized by the throat width $d_m$, the dimensionless
difference falls in relation to the nozzle size,
quantified by the ratio ${\Delta x_c\over d_m}$ shown in the right column.
In abolute numbers, $\Delta x_c$ grows with size (middle column),
until it actually drops for the largest nozzle.
Surprisingly, our atomistic simulations indicate that for a sufficiently large nozzle
the sonic horizon is situated right in the middle, with atomistic precision.

\begin{table}
	\begin{center}
		\begin{tabular}{|r|r|r|}
		\hline 
		$d_m$ & $\Delta x_c$ & ${\Delta x_c\over d_m}$ \\ 
		\hline 
		 3.90 & 3.78 & 0.97 \\ 
		 7.80 & 4.97 & 0.64 \\ 
		15.60 & 5.96 & 0.38 \\ 
		31.25 & 6.09 & 0.19 \\ 
		62.50 & 2.74 & 0.044 \\ 
		\hline 
		\end{tabular} 
	\end{center}
	\caption{Downstream shift $\Delta x_c$ of the sonic horizon with respect
          to the center position predicted by continuum fluid dynamics.
          Nozzle are characterized by the minimal width $d_m$. The right column
          shows the dimensionless difference in relation to nozzle size, ${\Delta x_c\over d_m}$.}
        \label{tab:AbsolutePosSonicHorizon}
\end{table}


\subsection{Phase diagram}

\begin{figure}[htbp]
	\begin{center}
	\includegraphics[width=1\columnwidth]{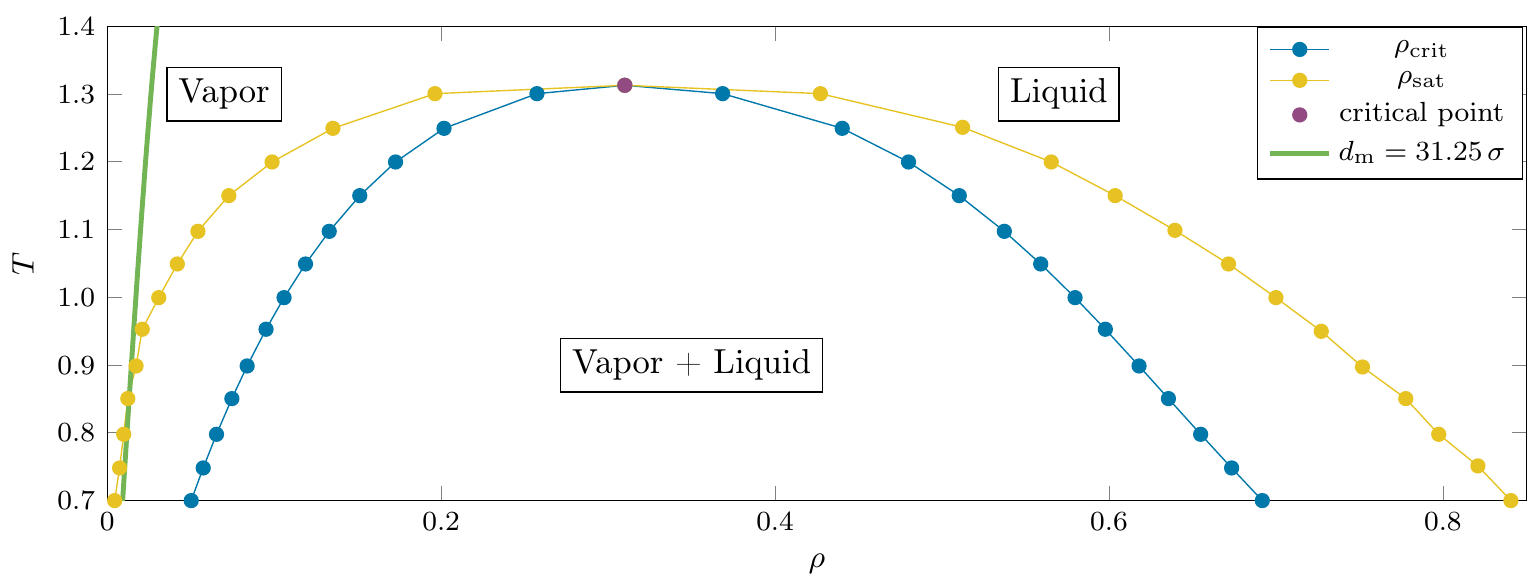}
	\caption{Density-temperature phase diagram. Shown are the saturation density (yellow),
          the critical density (blue) and the critical point (purple) from the Lennard-Jones
          equation of state \cite{johnson1993lennard}. For the nozzle with a throat width
          $d_m=31.25$ the path of temperature and density values are shown as green curve.}
	\label{PhaseDiagram}
	\end{center}
\end{figure}

Does the gas undergo a phase transition and condense into droplets at the end of the nozzle
as it cools upon expansion? Fig.~\ref{PhaseDiagram} shows the phase diagram of the LJ equation
of state in the $(T, \rho)$ plane as determined form Ref.~\cite{johnson1993lennard}.
The saturation density curve shown in yellow is associated with the phase transition,
but up to the critical density, shown as blue curve, a supersaturated vapor phase or
a superheated liquid phase is possible. This supersaturated and superheated phases are metastable.
The green curve in Fig.~\ref{PhaseDiagram} shows the path of density and temperature values,
shown in panels c) and i) of Fig.~\ref{picRasterAxis0p5}, of the gas expansion in the
nozzle with throat width $d_m=31.25$. Strictly speaking, only an adiabatically slow
evolution of a LJ fluid has a well-defined path in diagram Fig.~\ref{PhaseDiagram}, which shows
{\it equilibrium} phases. But plotting the state during expanding through the microscopic
nozzle in Fig.~\ref{PhaseDiagram} at least provides a qualitative description of the fluid
at a particular position in the nozzle. The path would extend to about $T=0.4$, but
the equation of state from ref.~\cite{johnson1993lennard} does not reach below $T=0.7$. We note
that the triple point, obtained from molecular simulations studies in Ref.\cite{ahmed2009solid}
lies at $T_{tr}=0.661$, below which the gas-liquid coexistence region becomes a gas-solid
coexistence region.

From the path traced by the expanding gas we see that the LJ fluid
starts in the gas phase in the inlet. As temperature and density fall upon expansion,
the fluid enters the gas-liquid coexistence region. In this region the fluid can remain in
a metastable supersaturated gas phase. Below the triple point, even the gas-solid coexistence
region is reached at the end of the nozzle. 

Our simulations show no evidence of a liquid or even a solid phase in our simulations, which would
appear as small liquid or solid clusters; the LJ particles remain unbound until
reaching the outlet region of the nozzle. Either the gas remains metastable or it is
to far out of local thermal equilibrium that the discussion in terms of the
phase diagram is meaningless. The anisotropy of the temperature discussed
in the previous section indicates that thermal equilibrium is not completely fulfilled.
The absence of nucleation of clusters is not a surprise, because there is simply not
enough time in a microscopic nozzle for nucleation under such dilute conditions
before the gas reaches the outlet.


\section{Microscopic Properties}
\label{sec:microscopic}

Molecular dynamics simulation allows to measure properties which are inaccessible
in a macroscopic continuum mechanical description. We already have seen in the previous
section the temperature is slightly anisotropic, which is inconsistent with local equilibrium.
In this section we take a closer look at quantities defined on an atomistic
level: the velocity probability distribution (in equilibrium the
Maxwell-Boltzmann distribution) and the velocity autocorrelation function.
Furthermore we study
the propagation of density waves by calculating the upstream and downstream time-correlations
of thermal density fluctuations of the stationary flow before, at, and after the sonic horizon.
The goal is to check if the sonic horizon, found in the previous section by
thermodynamic consideration, is also a well-defined boundary
for upstream information propagation on the microscopic level.

\subsection{Velocity Distribution}\label{sec:VelocityDistribution}

We have observed a temperature anisotropy, see panel c) and d) in Fig.\ref{picRasterAxis0p5}.
This raises the question whether the particle velocities even follow a Maxwell-Boltzmann
distribution. If the velocities are not Maxwell-Boltzmann distributed,
we do not have a well-defined kinetic temperature. This question is important for the
interpretation of the results, for example when we discussed the temperature drop during
expansion in the previous section. We now clarify whether it is meaningful to
talk about temperature in microscopic nozzles.

\begin{figure*}[htbp]
	\begin{center}
	\includegraphics[width=0.82\textwidth]{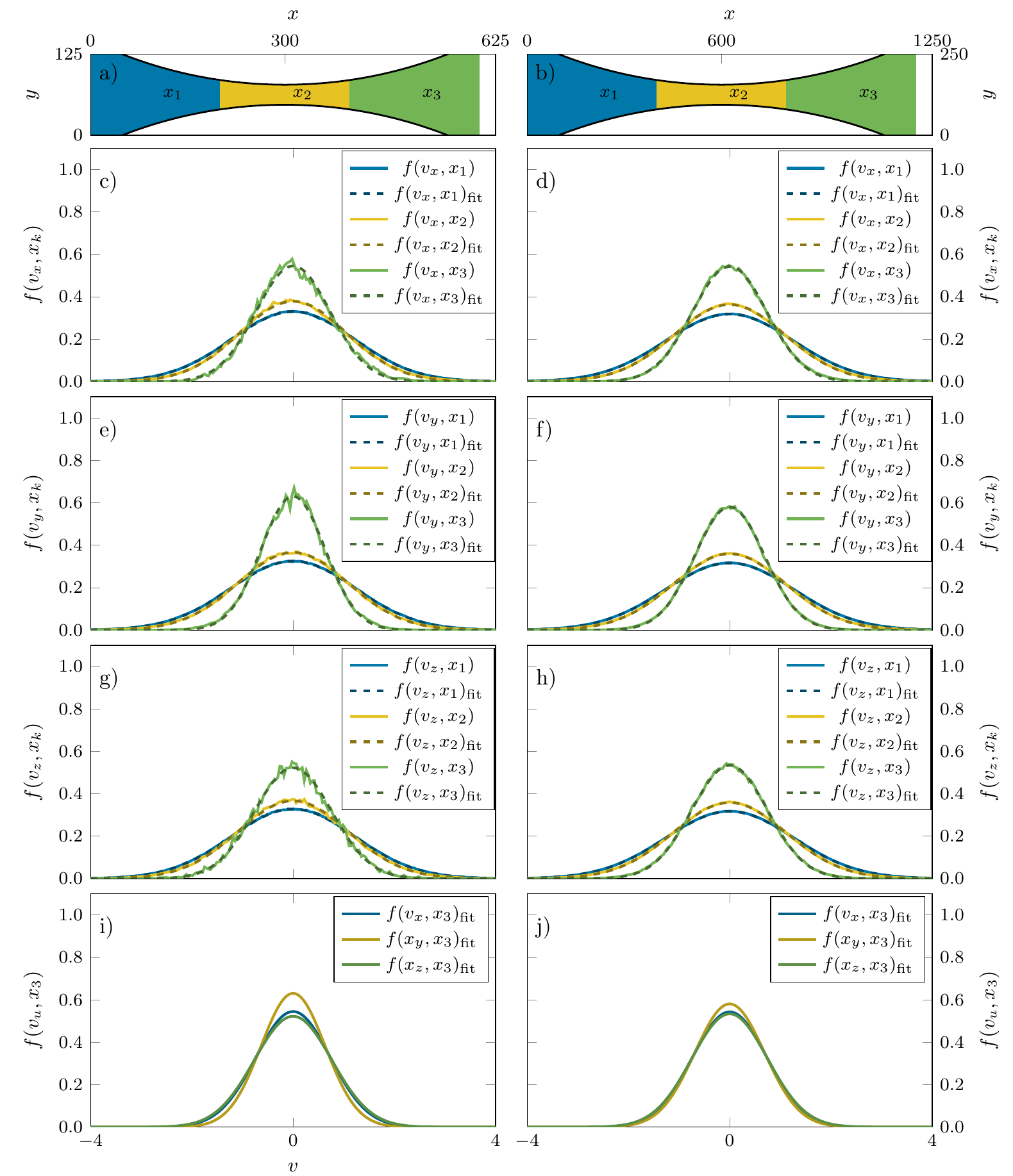}
	\caption{Thermal part of the particle velocity distribution $f(v_i,x_j)$ for two nozzle sizes with a
          throat width of $31.25\,\sigma$ and $62.5\,\sigma$ in the left and right column. respectively.
          Panels a) and b) show a schematic representation of those nozzles with the three
          regions $x_1$, $x_2$, and $x_3$ for which the velocity distributions are obtained from
          the MD simulations.
          Panels c) to h) show the velocity distribution of the components $v_x$, $v_y$ and $v_z$
          for the the different regions in the nozzle. Also shown are Gaussian fits (dashed lines).
          Panel i) and j) are comparing the fits to the three velocity components these fits
          in the $x_3$ region, in the diverging part of the nozzle.}
	\label{picVelDistribution}
	\end{center}
\end{figure*}

We calculate the velocity distribution for the two largest nozzles (see Fig.\ref{picRasterAxis0p5}).
shown in Fig.~\ref{picVelDistribution} by separately sampling the histograms
for the $x$, $y$, and $z$-components of the velocity, where we subtract the steady
flow velocity from the particle velocities, see appendix~\ref{sec:appD}.
Since the velocity distribution depends on the location $x$ in the nozzle,
the histograms are two-dimensional, which requires a lot of data to sample from. Therefore
we split $x$ into only three regions $x_1$, $x_2$ and $x_3$, depicted in the nozzle illustrations
at the top of Fig.~\ref{picVelDistribution}.

The velocity distributions $f(v_x,x_j)$ for the
$x$-component of the velocity are shown in panels c) and d) for the two respective nozzles,
each panel showing $f(v_x,x_j)$ for all three regions $x_j=x_1,x_2,x_3$ in blue, yellow, and
green. Of course, the distributions become more narrow for larger $x_j$, consistent with
a downstream drop of temperature in a Laval nozzle. We fit the histograms
with Gaussian functions, i.e. the Maxwell-Boltzmann distribution, also shown in the
panels. The corresponding results $f(v_y,x_j)$ and $f(v_z,x_j)$ for the other two velocity
directions are shown in panels e)--h).
It is evident that, apart from small statistical fluctuations, the Maxwell-Boltzmann
distribution is a good fit in all cases.
Thus the notion of temperature in these microscopic non-equilibrium systems makes sense.

The width of the velocity distributions (i.e. the temperature) is not quite the same
in the three directions, however, in particular in region $x_3$, the diverging part of the nozzle.
In order to see this better, we compare the fits to $f(v_i,x_3)$ for $i=x,y,z$ in panels i) and j).
The distribution of the $y$-component of the velocity is narrower than the other two directions.
In other words the temperature according to $v_y$ is lower, thus the temperature
is not isotropic. This means there is insufficient equilibration between the three
translational degrees of freedoms. The effect is more pronounced for the smaller nozzle
because particles undergo fewer collisions before
they exit the nozzle, as quantified by the larger Knudsen number, see Fig.\ref{picRasterAxis0p5}.

The spatial binning into just three region $x_j$ is rather coarse-grained as it neglects
the temperature variation within a region. With more simulation
data a finer spatial resolution would be possible, however we feel that the presented
results are convincing enough that the thermal kinetic energy can be well-characterized
by a temperature, albeit slighly different in each direction.

\subsection{Velocity Autocorrelation Function}\label{sec:VACF}

The velocity auto-correlation function, \gls{vacf}, quantifies the ``memory'' of particles
about their velocity. The VACF is defined as
\begin{equation}\label{eqVACF1}
	\mathrm{VACF}(\tau)=\big\langle\qv_p(t)\cdot \qv_p(t+\tau)\big\rangle_{t,p}
\end{equation}
with $\qv_p(t)$ the velocity of particle $p$ at time $t$. $\langle\dots\rangle_{t,p}$ denotes
an average over time and over all particles. An ideal, i.e.\,non-interacting particle
has eternal memory, $\mathrm{VACF}_u(\tau)={\rm const}$. But due to interactions with the other
particles, $\mathrm{VACF}_u(\tau)\to 0$ within microscopically short times.

\begin{figure*}
	\begin{center}
		\includegraphics[width=0.8\textwidth]{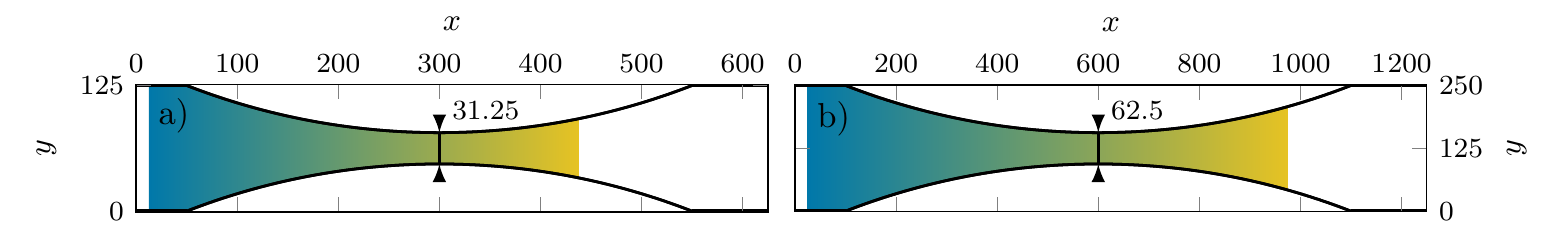}
		\includegraphics[width=0.36\textwidth]{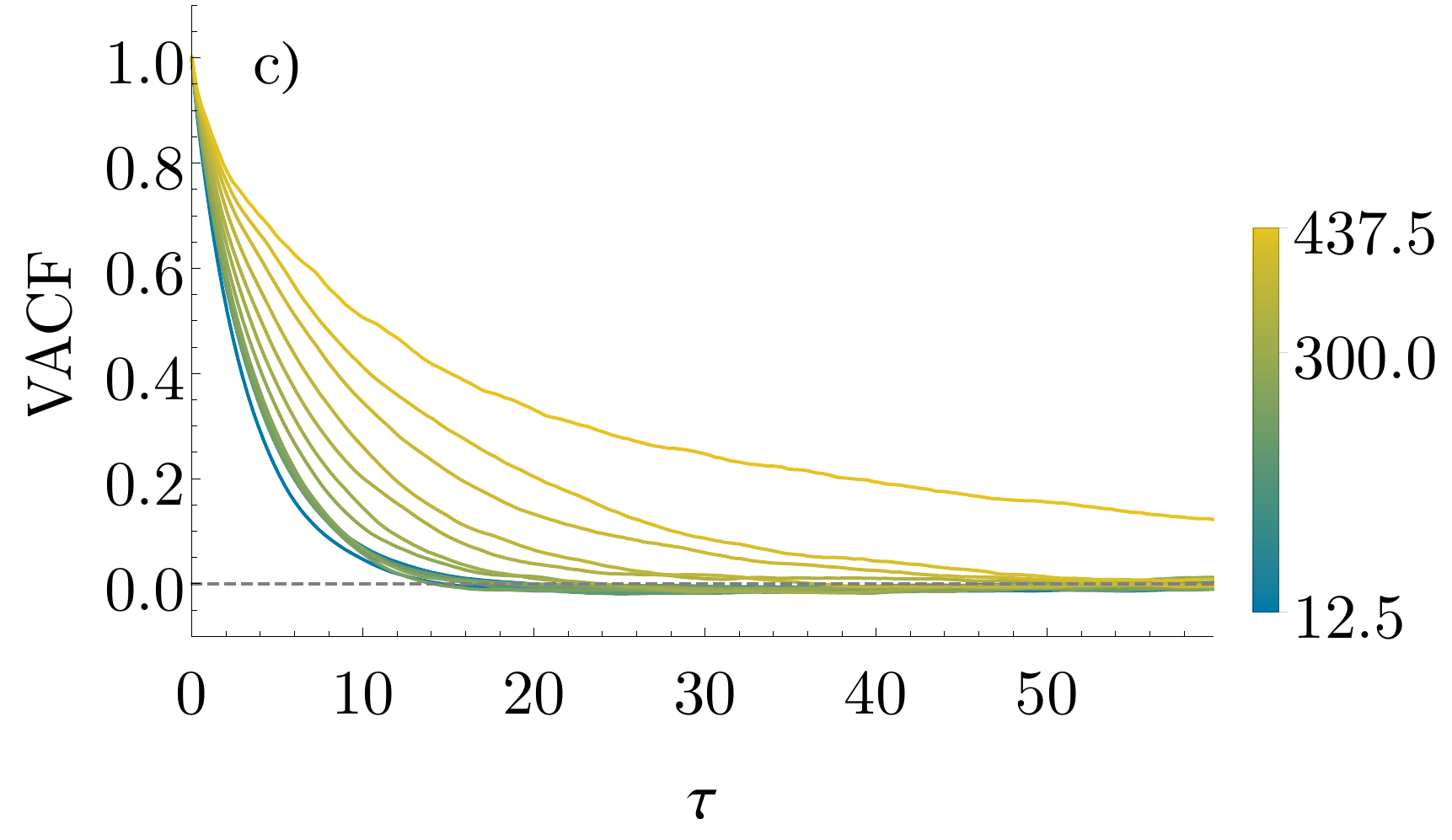}
		\includegraphics[width=0.36\textwidth]{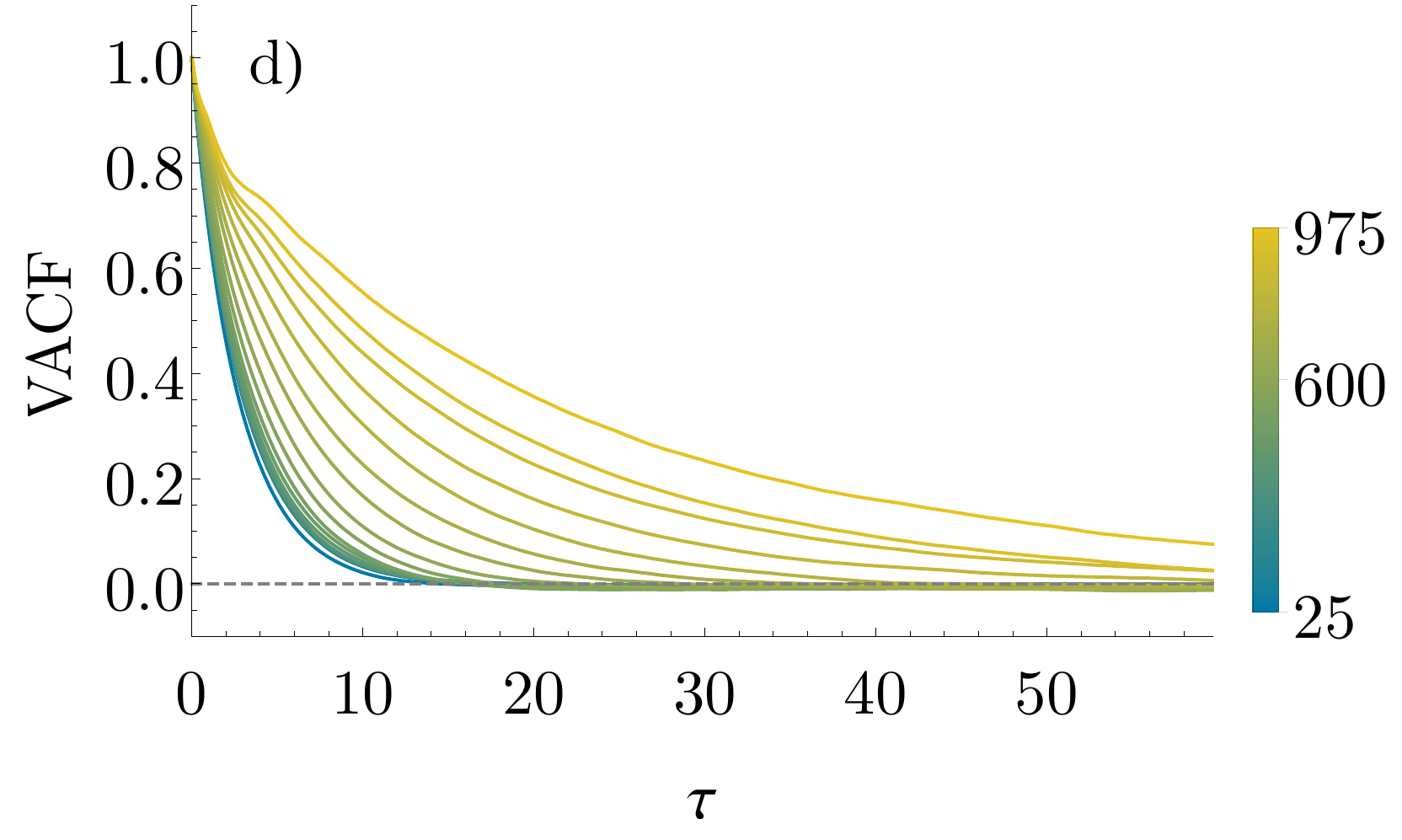}
		\includegraphics[width=0.36\textwidth]{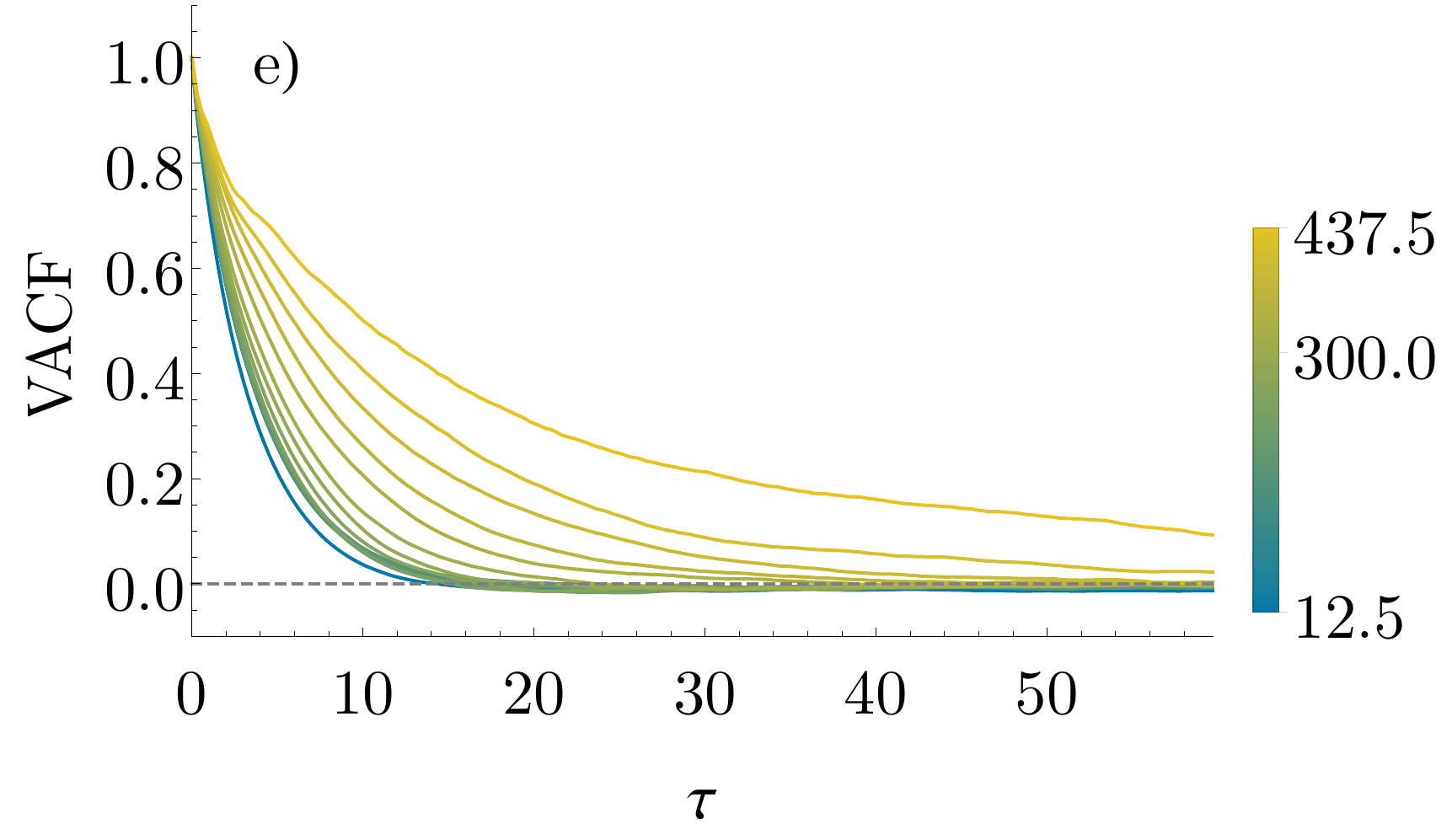}
		\includegraphics[width=0.36\textwidth]{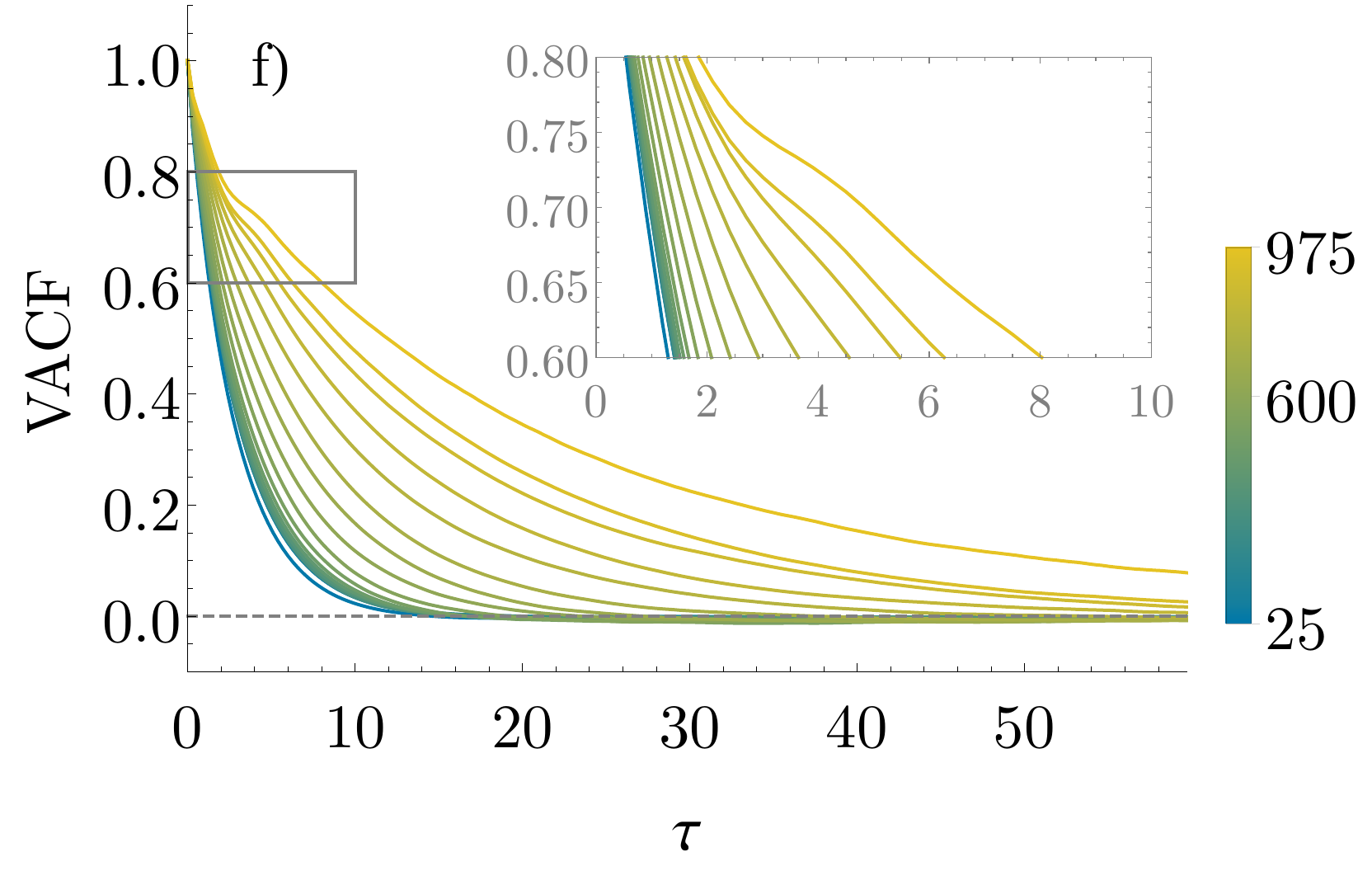}
	\end{center}
	\caption{Normalized velocity auto-correlation function $\mathrm{VACF}(x,\tau)$,
          eq.~(\ref{eqVACF1a})),
          along the nozzle with color coded $x$-position. Panels a) and b) show the shape and size
          of two nozzles, indicating the color scale for $x$ in the panels below.
          Panels c) and d) show the \gls{vacf} for nozzles with a distance $z_{max}=43.1\,\sigma$
          between the periodic boundaries in $z$-direction. Panels e) and f) show the same
          for $z_{max}=86.2\,\sigma$. The inset of panel f) shows a close-up of
          the shoulder around $\tau=4$, discussed in the text.}
	\label{pic:VelvAutoCorrLin}
\end{figure*}

In the case of stationary flow, we need to subtract the flow velocity from particle velocities
in eq~(\ref{eqVACF1}). Furthermore, the VACF will depend on the $x$-coordinate in the nozzle.
Therefore we generalize eq.~(\ref{eqVACF1}) to
a form which is suitable for stationary flow in a nozzle that depends on $x$ and is not biased
by the flow velocity. We also normalize the VACF such that is is unity at $\tau=0$:
\begin{multline}
  \mathrm{VACF}(x,\tau)
  ={\big\langle\Delta\qv_{p}(t)\cdot\Delta\qv_{p}(t+\tau)\,\delta(x-x_p(t))\big\rangle_{t,p} \over
    \big\langle \Delta\qv_{p}(t)^2\,\delta(x-x_p(t))\big\rangle }
  \label{eqVACF1a}
\end{multline}
where $\Delta\qv_p(t)\equiv \qv_p(t)-\qv(x_p(t))$ is the thermal part of the velocity,
after subtraction of the flow velocity $\qv$ at the particle coordinate $x_p(t)$. Note that
we define $\mathrm{VACF}(x,\tau)$ such that the spatial coordinate $x$ coincides with the
starting point $x_p(t)$ at time $t$ of the time correlation; at the final time
$t+\tau$, the particle has moved to $x_p(t+\tau)$ downstream.
When we sample (\ref{eqVACF1a}) with a MD simulation, the coordinate $x$
and the correlation time $\tau$ are discretized, and $\delta(x-x_p(t))$ is replaced by
binning a histogram in the usual fashion, see the appendix.

Fig.~\ref{pic:VelvAutoCorrLin} shows the \gls{vacf} for various positions
$x$ in the nozzle. The calculations were done for two different nozzle sizes
(left and right panels). The \gls{vacf}s cannot be shown for $x$ all the way to the
end of the nozzles because particles leave the simulation before the velocity
correlation can be evaluated. For example, if a particle in the smaller of the two nozzles
in Fig.~\ref{pic:VelvAutoCorrLin} is located at $x=437$ at $\tau=0$ it will have moved
with the flow on average to $x=537$ at $\tau=50$, where the outlet region starts and particles
are removed from the simulation. For $x$ close to the outlet,
the VACF would be biased because the average in eq.~(\ref{eqVACF1a}) would contain
only particles which happen to travel slow, e.g.\ slower than the flow average.

The \gls{vacf} decays monotonously for all $x$ (in fact,
the \gls{vacf} for only the $y$-component of the velocity (not shown) slightly overshoots to a
negative correlations in the divergent part of the nozzle, which is a trivial
effect of wall collisions). The decay is slower further downstream
because the density drops.
Towards the ends of the nozzles, the mean free path becomes large, see
Fig.~\ref{picRasterAxis0p5}, reaching the length $z_{max}$ of the simulation box in
$z$-direction, where periodic boundary conditions are applied.
We demonstrate that the finite size bias in $z$-direction is negligible by comparing the
VACFs for different choices of $z_{max}$. If $z_{max}$ were too small,
two particles might scatter at each other more than once due to the periodic
boundaries, which would lead to a spurious oscillation in the VACF. 
Panels e) and f) in fig.~\ref{pic:VelvAutoCorrLin} show $\mathrm{VACF}(x,\tau)$
for $z_{max}=86.2\sigma$, twice as large as in panels c) and d), corresponding to
twice as many particles. Apart from the smaller statistical noise for larger $z_{max}$,
the VACFs for $z_{max}=43.1\sigma$ and $z_{max}=86.2\sigma$ are identical.
This confirms that $z_{max}=43.1\sigma$ is large enough to obtain reliable results.

An interesting feature in the VACF for both nozzle sizes shown in Fig.~\ref{pic:VelvAutoCorrLin}
is a small shoulder around $\tau\approx 4$ in the divergent part, i.e.\, a small
additional velocity correlation. The inset in panel f) of Fig.~\ref{pic:VelvAutoCorrLin}
shows a close-up of the shoulder.
Since this happens only at the low density in the divergent part of the nozzle, where the
three-body collisions rate is low, the shoulder can be expected to be a two-body effect.
It is consistent with pairs of particles orbiting around each other a few times.
We test this conjecture by estimating the orbit period of two bound atoms in
thermal equilibrium. The orbit speed $v$ shall be determined by the temperature $T$.
We further assume a circular stable orbit with diameter $d$.
The orbiting particles have two rotational degrees of freedom
but also two times the mass of a single particle:\\
\begin{equation}\label{eq:OrbitTime1}
 \frac{1}{2} k_B T = \frac{1}{2} m v^2.
\end{equation}
The centrifugal force $F_c$ and the attractive \gls{lj} force $F_{LJ}$ must be balanced,
\begin{equation}\label{eq:OrbitTime2}
	F_c+F_{\rm LJ} = m \frac{2 v^2}{d}
	- 4 \epsilon m\left(-\frac{12}{d^{13}} +\frac{6}{d^7} \right)=0,
\end{equation}
The orbit period $t_{rot}$ can now be calculated from 
eq.~(\ref{eq:OrbitTime1}) and eq.~(\ref{eq:OrbitTime2})
\begin{equation}
	t_{\rm rot}=\pi \frac{d}{v}
	=\pi\left(
		\frac{6 \epsilon m^4\pm\sqrt{36\epsilon^2 - 24 k_B T/m}}{k_B^4 T^4}
	 \right)^{1/6}\label{eq:OrbitTime3}.
\end{equation}
which expresses $t_{\rm rot}$ as function of the temperature. When we plug in a typical
temperature towards the end of the nozzles of $T\approx 0.5$, we obtain an orbit time
$t_{\rm rot}\approx 5$, which is similar to the time when the shoulder
in the VACF appears, see Fig.~\ref{pic:VelvAutoCorrLin}.
This does not mean that bound dimers form in the supercooled flow near the exit of
the nozzle, which requires three-body collisions. But the estimate based on bound states
is applicable also to spiral-shaped scattering processes where two particles orbit each
other. The good agreement between the $t_{\rm rot}$ and the shoulder indicates that
such scattering processes occur, and may be a seeding event for the nucleation of van der Waals
clusters and condensation in larger nozzles.

\subsection{Density fluctuation correlations and the sonic horizon}

The calculation of the speed of sound $c$ according to eq.(\ref{equDefSpeedOfSound}),
using the equation of state
from Ref.\cite{johnson1993lennard}, assumes local thermal equilibrium. However, the anisotropy of
the temperature, see Fig.~\ref{picRasterAxis0p5}, shows that not all degrees of freedom
are in local equilibrium during the fast expansion through a microscopic nozzle.
Therefore, locating the sonic horizon may be biased by non-equilibrium effects.
It's not even clear if a sonic horizon, the definition of which is based on macroscopic
fluid dynamics, is microscopically well-defined. While 
the thermal velocities of the atoms follow Maxwell-Boltzmann distributions,
there are always particles in the tails of the distribution
that travel upstream even after the sonic horizon. So maybe information can travel upstream
on the microscopic scale of our nozzles, negating the existence of a sonic horizon.

\begin{figure*}[htbp]
  \begin{center}
  \includegraphics[width=0.65\textwidth]{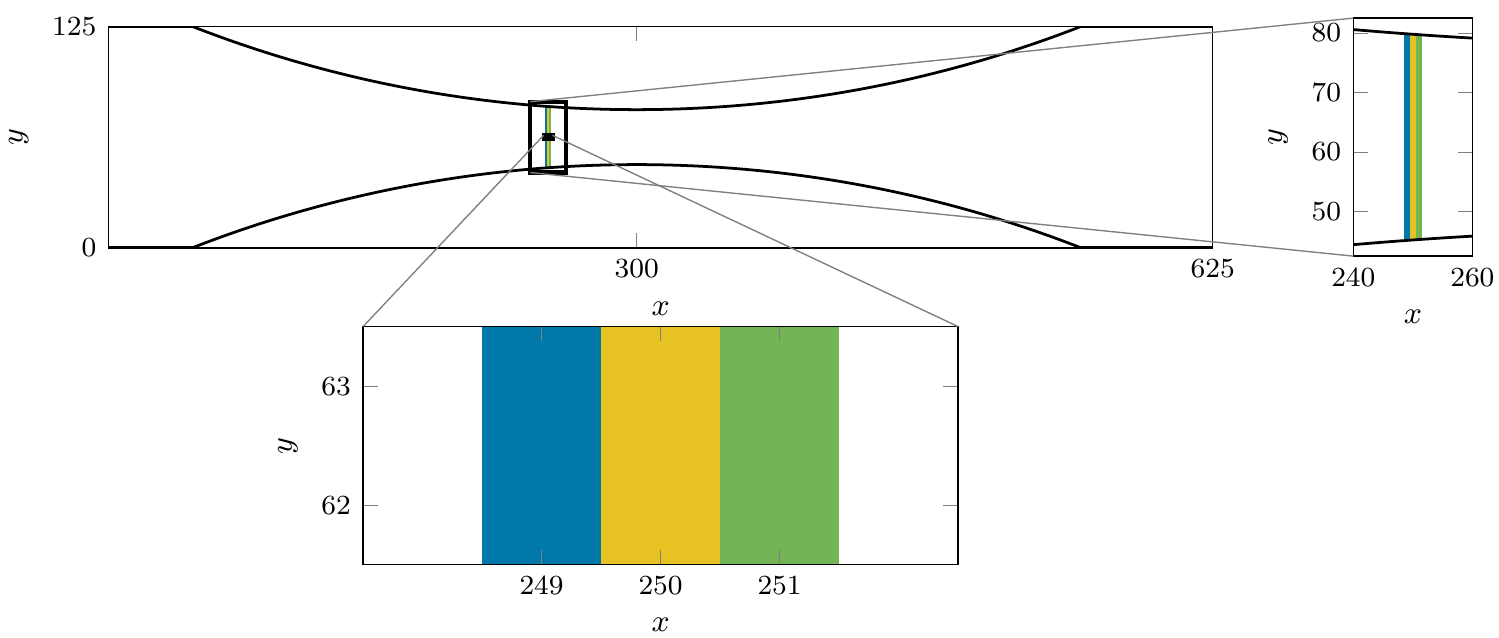}
  \includegraphics[width=0.82\textwidth]{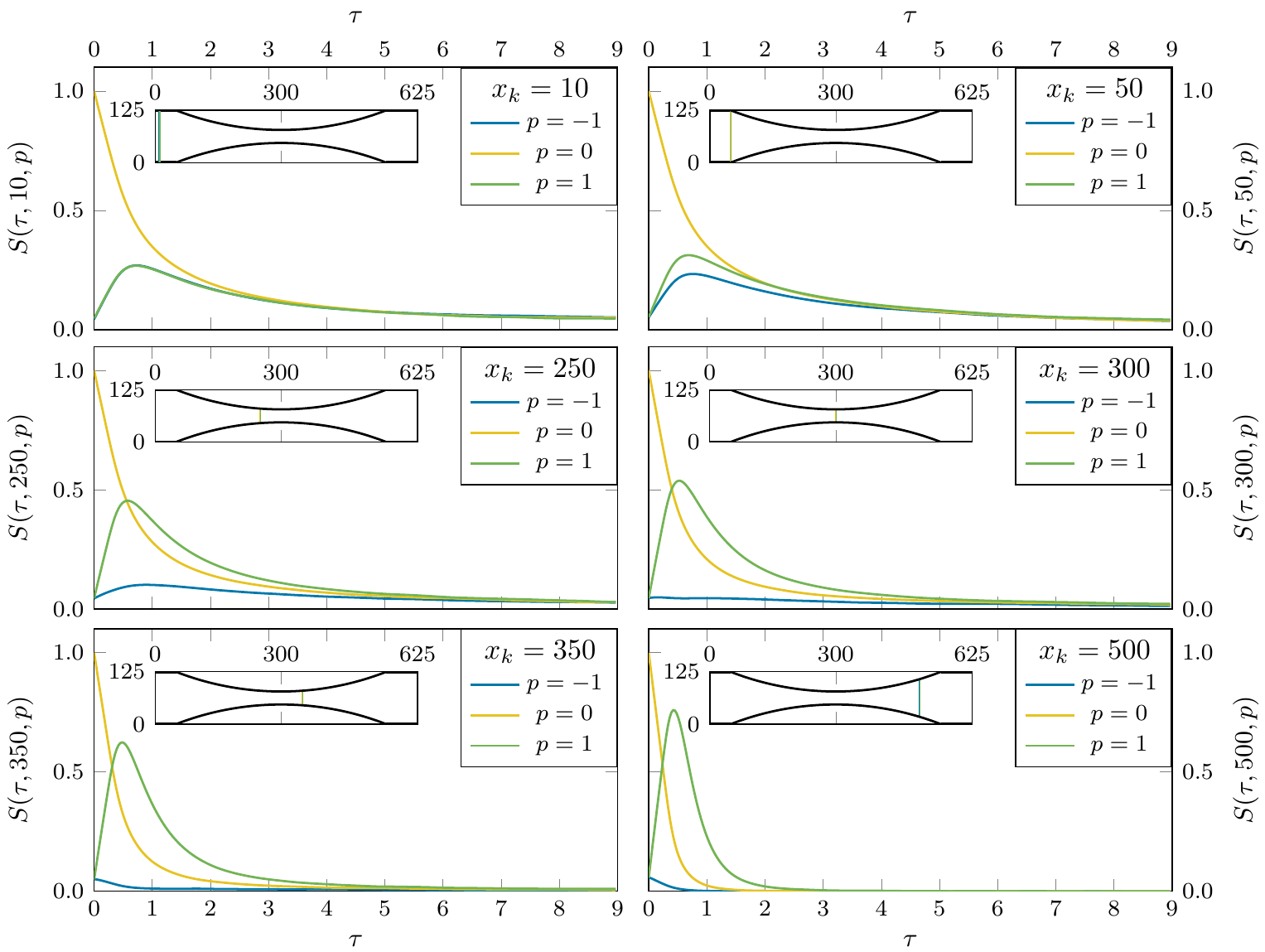}
  \end{center}
  \caption{Density fluctuation correlations $S(\tau,x,\delta x)$,
	eq.~(\ref{eq:DensityFluctuationCorr1}). Panels a) to f) show the self correlation
	$S(\tau,x,0)$ in yellow, a backward correlation $S(\tau,x,-\sigma)$ in blue and
	a forward correlation $S(\tau,x,\sigma)$ in green for different positions
	$x$ in the nozzle as given in the insets. The illustration at the top
	shows the density bins used for calculating $S(\tau,x,\delta x)$:
	$S(\tau,x,0)$ is obtained by correlating the yellow bin with itself,
	$S(\tau,x,\sigma)$ or $S(\tau,x,-\sigma)$ are obtained by
	correlating the yellow bin with the green or blue bin, respectively.}
  \label{picDensityFlucCorrS0p5Sig1}
\end{figure*}

The \gls{md} methods provides the microscopic tools to answer this question
by calculating spacetime correlations of density fluctuations: 
if density fluctuations propagate upstream even in the divergent part of the nozzle,
there is no sonic horizon.
We quantify the density fluctuation correlations before, at, and
after the sonic horizon predicted from the calculation of the speed of sound.
The instantaneous density $\rho(x,t)$ at position $x$ and time $t$ is
evaluated according to eq.~(\ref{eq:Densitytime}).
The density fluctuation, i.e.\ the random deviation at time $t$ from the average density
at position $x$, is obtained by subtracting the time-averaged density (shown
in Figs.~\ref{picRasterAxis0p0312}, \ref{picRasterAxis0p1248}, and~\ref{picRasterAxis0p5})
from $\rho(x,t)$, $\Delta\rho(x,t)= \rho(x,t)-\big\langle \rho(x,t)\big\rangle_t$.
Note that fluctuations of the density depend also on $y$ and $z$, but we are interested
in the fluctuations
relative the the sonic horizon, and thus fluctuations between different positions $x$
in the nozzle. The correlation between a density fluctuation at $x$ and $t$ and a density
fluctuation at $x+\delta x$ and $t+\tau$ is given by the time average
\begin{equation}\label{eq:DensityFluctuationCorr1}
  S(\tau,x,\delta x) =
  \frac{\big\langle \Delta\rho(x,t)\, \Delta\rho(x+\delta x,t+\tau) \big\rangle_t}
       {\big\langle \Delta\rho(x,t)\, \Delta\rho(x,t) \big\rangle_t}
\end{equation}
$S$ is normalized such that it is unity for zero spatial and temporal shifts, $S(0,x,0)=1$.

In Fig.\,\ref{picDensityFlucCorrS0p5Sig1} we show the density fluctuation
correlations $S(\tau,x,\delta x)$ in a nozzle with throat width $31.25\sigma$,
evaluated at 6 different positions $x$ in the nozzle and for three relative position offsets
$\delta x=p\,\sigma$ with $p\in\{-1,0,1\}$. The position $x$ in the nozzle is indicated
in an inset in each panel. The density binning, with bin size $\sigma$, is illustrated at the top of
Fig.~\ref{picDensityFlucCorrS0p5Sig1},
which shows three adjacent bins at $x$, $x+\sigma$ and $x-\sigma$,
corresponding to $p/\sigma=-1,0,1$ in the figure labels.

The self correlation $S(\tau,x,0)$ (yellow curves), correlating only the temporal
decay of the density correlations at $x$, is mainly influenced by the flow velocity
and decays faster for higher flow velocities because density fluctuation are transported
away more quickly.

The upstream correlations $S(\tau,x,-\sigma)$ (blue curves) and the downstream
correlations $S(\tau,x,\sigma)$ (green curves) are more interesting. Both correlations
are small at zero delay time $\tau=0$, because a density fluctuation
at $x$ needs some time to disperse to neighboring density bins.  At position $x=10$, where the flow speed
is still small, there is no noticable difference between upstream and downstream
correlation. For larger $x$, hence for larger flow speed, the forward correlation
increases and the backward correlation decreases, because the density fluctuation
disperses with the flow or against the flow, respectively.

According to the local speed of sound calculated in the previous section,
see table~\ref{tab:AbsolutePosSonicHorizon}, there is
a sonic horizon at $x=306$ for the nozzle size in Fig.~\ref{picDensityFlucCorrS0p5Sig1}.
Indeed, for $x=300$, the backward correlation has no peak anymore, but decreases
monotonously from a small non-zero value at $\tau=0$. For even larger $x$,
the upstream correlation decays more rapidly, yet it never completely vanishes at
$t=0$.
The reason for this apparent contradiction to the existence of a sonic horizon is
that the distance between bins and the width of the bins are both $\sigma$.
The finite value at $\tau=0$ is an artifact caused by the density bins being directly
adjacent to each other, see the illustration in Fig.\,\ref{picDensityFlucCorrS0p5Sig1}:
a density fluctuation at $x$ will immediately have an effect on the adjacent bins
at $x+\sigma$ and $x-\sigma$ since they share a boundary.

\begin{figure*}[htbp]
	\begin{center}
		\includegraphics[width=0.82\textwidth]{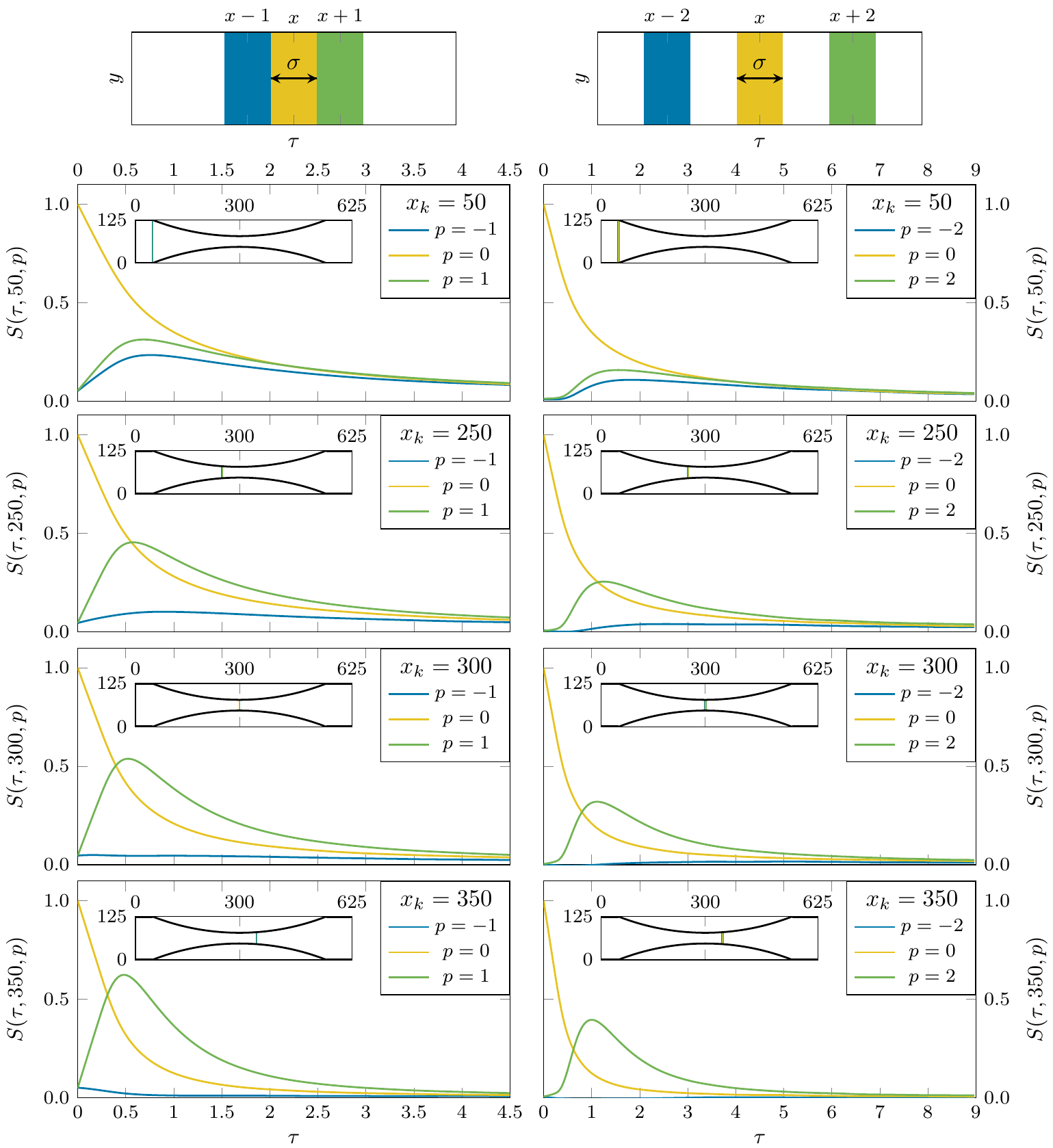}
	\end{center}
	\caption{Comparison of density fluctuation correlation $S(\tau,x,p\sigma)$ for different
	 offsets, $p\in\{-1,0,1\}$ (left panels) and $p\in\{-2,0,2\}$ (right panels).
	 At the top the respective binning is illustrated. In the insets the reference
	 position $x$ is indicated. The sonic horizon is situated slightly downstream of
	 the nozzle throat ($x=300$) at $x=306$, according to the thermodynamic
	 calculation of the local speed of sound.}
	\label{picDensityFluctuationSig2}
\end{figure*}

In order to remove this bias, we also calculated the correlations with offsets
$\delta x=\pm 2\sigma$, $S(\tau,x,2\sigma)$ and $S(\tau,x,-2\sigma)$, such that
the upstream and downstream bins do not share a boundary with the bin at $x$.
In Fig.~\ref{picDensityFluctuationSig2} we compare the two choices of offsets.
The left panels are take from Fig.~\ref{picDensityFlucCorrS0p5Sig1} where
$\delta x\in\{-\sigma,0,\sigma\}$; the right panels show $S(\tau,x,\delta x)$
with $\delta x\in\{-2\sigma,0,2\sigma\}$, with a twice as large $\tau$ range,
because density fluctuations have to travel twice as far.
The upstream and downstream correlations now vanish for zero time delay $\tau=0$.
The upstream correlation $S(\tau,x,-2\sigma)$ right at the throat
at $x=300\sigma$ is very small but does not quite vanish, which is consistent with
a location of the sonic horizon predicted at $x=306\sigma$ according to the
speed of sound. Further downstream at $x=350\sigma$, however,
$S(\tau,x,-2\sigma)$ indeed vanishes within the error bars. This means that information
about density fluctations cannot travel backwards beyond the sonic horizon even on the microscopic
scale of just a distance of $2\sigma$. A microscopic Laval nozzle does have a sonic horizon.

\begin{figure*}[htbp]
	\begin{center}
		\includegraphics[width=0.82\textwidth]{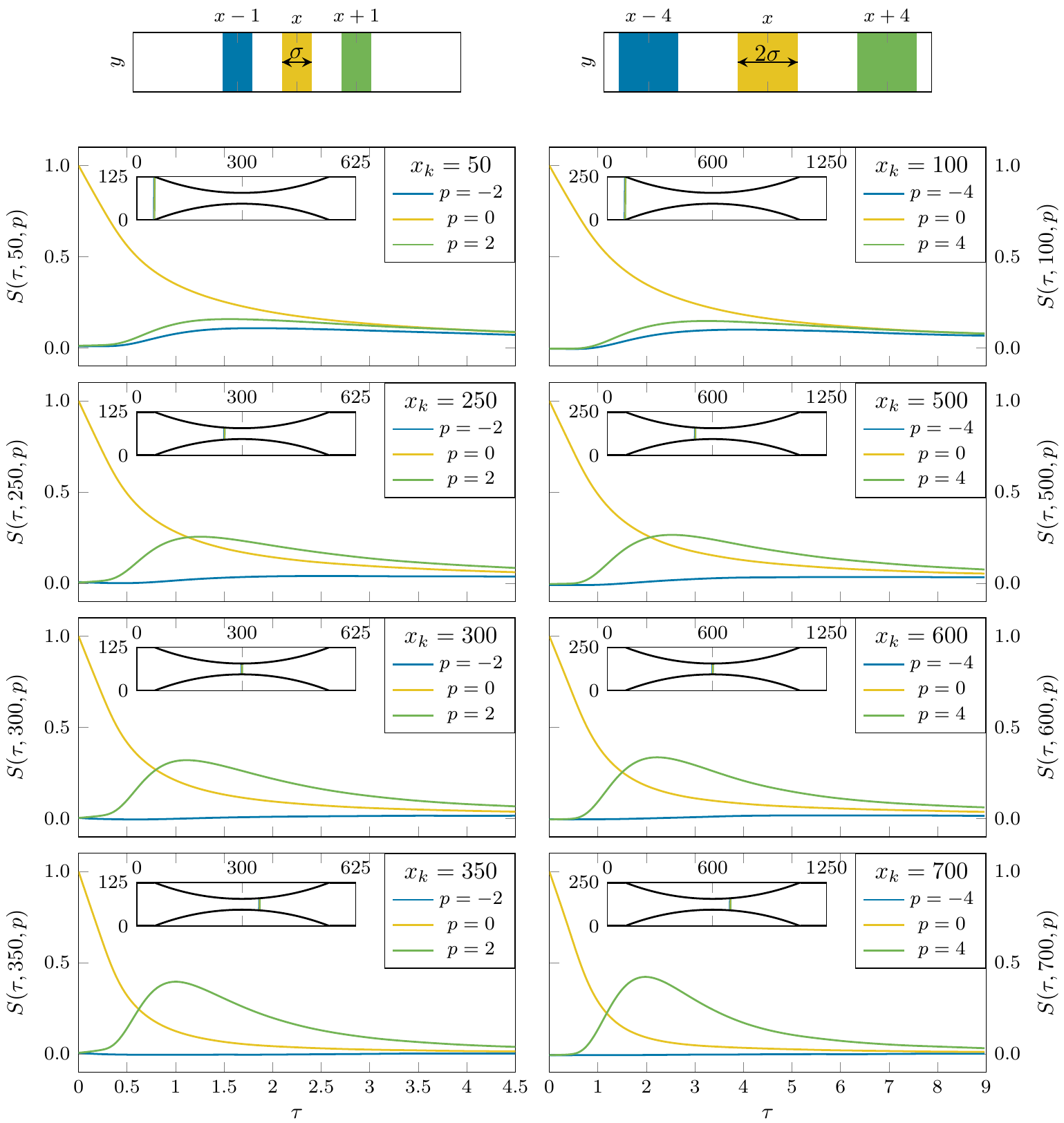}
	\end{center}
	\caption{Comparison of the density fluctuation correlations $S(\tau,x,\delta x)$ for
	two nozzle with throat width $d=31.25\,\sigma$ (left panels) and $d=62.5\,\sigma$
	(right panels),	respectively. All lengths are scaled by two for the larger nozzle,
	such that we compare the correlations for equal relative positions.
	At the top the density bin spacing is illustrated and the insets show the positions $x$.}
		\label{picDensitySizeComp}
\end{figure*}

We also calculated the density fluctuation correlations for a nozzle twice as large
(length $L=1250\,\sigma$ and throat width $d=62.5\sigma$).
Fig.~\ref{picDensitySizeComp} compares the correponding results with those shown
in Fig.~\ref{picDensityFluctuationSig2}. For the comparison, we
scaled all lengths by two: the bins are $2\,\sigma$ wide, separated by $4\,\sigma$,
see illustration at the top of Fig.\,\ref{picDensitySizeComp}. We compare
$S(\tau,x,\delta x)$ of the smaller nozzle with $S(2\tau,2x,2\delta x)$ of the larger one,
i.e.\ at the same relative positions with the same relative upstream
and downstream offset, and showing twice the time window for the larger nozzle.
According to the speed of sound, the sonic horizon for the larger nozzle is located
at $x=603\,\sigma$ (see table~\ref{tab:AbsolutePosSonicHorizon}),
very close to the throat at $x=600\,\sigma$. The comparison in Fig.~\ref{picDensitySizeComp} shows
that the density fluctuation correlations are very similar for equal relative positions
for both nozzles. Also for the larger nozzle, the correlations are very small at the throat.
Further downstream at $x=350\sigma$ and $x=700\sigma$, respectively, both nozzles exhibit no
upstream correlations.

Our calculations confirm
that the thermodynamic determination of a sonic horizon, based on the equation of state,
is valid, although the anisotropy of the temperature indicates that the rapid expansion
through the nozzles hinder complete local thermal equilibrium.
The location of the sonic horizon is consistent with the vanishing of upstream time correlations
of density fluctuations.
The existence of a microscopically narrow sonic horizon is a non-trivial result,
considering the large estimated Knudsen numbers.

\section{Conclusion}

We studied the expansion of a gas of Lennard-Jones particles and its transition from
subsonic to supersonic flow through microscopic Laval slit nozzles into vacuum. Our goal was to assess to
what extent Laval nozzles with throat widths down to the scale of a few atom diameters still follow the
same mechanisms as macroscopic nozzles where, given a sufficiently low outlet pressure,
the gas flow becomes supersonic in the nozzle throat.  For our study we used non-equilibrium
molecular dynamics (MD) simulations. MD is computationally demanding but makes the fewest
approximations. We considered idealized nozzles with atomically flat surfaces with perfect slip to
avoid boundary layer effects.

We introduced three thermodynamic regions for the non-equilibrium molecular dynamic simulation:
an inlet region, the nozzle region and the outlet region. In the inlet and outlet region,
particle insertions and deletions are realized by grand canonical Monte Carlo sampling~\cite{heffelfingerJCP94}.
After equilibration this allows to study stationary flows.

We obtained the thermodynamic state variables temperature, density, flow velocity, and pressure and
their spatial dependence, as well as the Knudsen number, Mach number,
velocity auto-correlation, and velocity distribution of the gas for nozzles
of different sizes. We found a well-defined sonic horizon, i.e.\,the surface where the flow becomes
supersonic, and analyzed it via spacetime correlations of density fluctuations.
We studied how the expansion dynamics depend on the nozzle size. Lower temperatures and
correspondingly higher velocities and Mach numbers of the expanding gas are reached for larger nozzles,
converging to predictions for isentropic expansion of an ideal gas continuum.

With non-equilibrium molecular dynamics we can observe phenomena which cannot be studied
in continuum fluid dynamics, which assumes local thermodynamic equilibrium.
We found that this assumption is violated for microscopic nozzles. The kinetic energy
in the three translational degrees of freedom cannot equilibrate completely and is slightly
different for each individual translational degree of freedom.
The velocity components are still Maxwell-Boltzmann distributed, with a different width
for each direction, which corresponds to an anisoptropic temperature.

The phase of the LJ fluid in the inlet is in a vapor phase, but upon expansion
through the nozzle becomes supersaturated. At the end of the nozzle it is in the
vapor-solid coexistence phase. Indeed, in the velocity auto-correlation function,
\gls{vacf}, we see indications of metastable pairs of particles. Since the expanding gas does
not reach equilibrium in our microscopic nozzles, no clusters are formed.
Cluster formation could be studied by enlarging
the simulation and including the low density region after the nozzle, giving the fluid
enough time to equilibrate.

The investigation of the sonic horizon with the help of spacetime-dependent correlations of
density fluctuations showed that the position of the sonic horizon obtained from calculating
the local speed of sound
matches the position where density correlations practically cannot propagate against
the flow. A microscopic distance on the order to the LJ particle size $\sigma$ is already
enough to completely suppress the backward correlations. The vanishing of backward time correlations
does of course not happen abruptly at the sonic horizon, instead the backward correlations decrease
gradually with the increasing flow velocity toward the sonic horizon. At the same time the forward
correlations increase with the flow velocity. For larger microscopic nozzles, the simple macroscopic
description relating the cross section to the Mach number is quite accurate. For smaller nozzles
the position of the sonic horizon is shifted downstream.

In future work, it will be interesting to study nozzles with rough walls.
The gas expansion through microscopic nozzle will be
strongly affected by the boundary layer near the walls. Another topic of practical interest is the
co-expansion of a carrier noble gas seeded with molecules to investigate
the cooling efficiency of rotational and vibrational degrees of freedom of the molecules. This
models the cooling of molecules for molecular beam spectroscopy. We note that nozzles for
molecular beam spectroscopy are significantly larger than those studied here, with nozzle diameters
of the order of tens of $\mu m$, instead of tenths of $nm$.
Increasing the outlet region
will allow to study not only the condensation of the gas into clusters, but
also the effect of a finite exit pressure on the position of the
sonic horizon~\cite{saadatiAST15}.

We acknowledge inspiring discussions with Stefan Pirker.

\begin{appendix}

\section{Density calculation}\label{sec:appA}

\begin{figure}[htbp]
	\begin{center}
		\includegraphics[width=1\columnwidth]{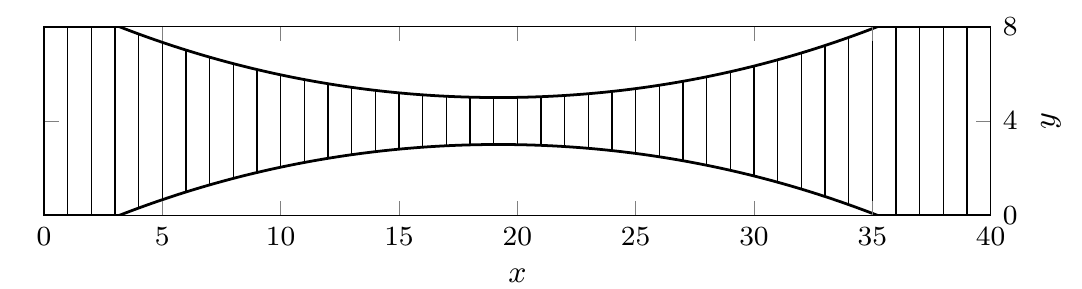}
	\end{center}
	\caption{Bin volumes of width $\sigma$ used for calculating the density $\rho(x)$.}
		\label{DensitySlices2}
\end{figure}

The density $\rho(x)$ as function of position $x$ in the nozzle is calculated by
binning the $x$-coordinate of all particles. Since we are interested in stationary
flow situations, we can take time averages of the number of particles in the bin
of volume $V_\mathrm{bin}(x)$. The binning volumes are slices, usually of
thickness $\sigma$, which are centered at $x$, as illustrated in Fig.~\ref{DensitySlices2}.
This average can be written as
\begin{equation}\label{eq:Density}
  \rho(x)=\Bigg<\frac{1}{V_\mathrm{bin}(x)}\sum_{i:p_i \in V_\mathrm{bin}(x)} 1\Bigg>_t
  \equiv \big< 1 \big>_{t,V_\mathrm{bin}(x)}
\end{equation}
with the sum counting all particles $p_i$ in the volume of bin $V_\mathrm{bin}(x)$, and
the bracket denoting the time average.
For calculations of spacetime density correlations we need the instantaneous density
at $x$ at time $t$, which we obtain by omitting the time average in eq.~(\ref{eq:Density})
\begin{equation}\label{eq:Densitytime}
  \rho(x,t)=\frac{1}{V_\mathrm{bin}(x)}\sum_{i:p_i \in V_\mathrm{bin}(x)} 1
\end{equation}

The determination of $V_\mathrm{bin}(x)$ is not trivial, since the wall is not a well-defined
hard boundary, but realized by the LJ potential~(\ref{equLJWall}).
Choosing $z=0$ in eq.~(\ref{equLJWall}) for the volume calculation would overestimate the real
volume effectively available for the particles, because it neglects the thickness
of the ``skin'' due to the finite value of $\sigma$. We determined that $z=0.8\,\sigma$ is the most
suitable choice in the following way: we simulated a small nozzle (the size depicted
in Fig.~\ref{DensitySlices2}) with a constriction so narrow that almost no particle pass
through in the course of a simulation. The wall position $z$, and hence the
effective volume $V_\mathrm{bin}(x)$, is determined such that the density $\rho(x)$
in the left half of the nozzle,
obtained from (\ref{eq:Density}), is constant as expected for an equilibrium
simulation in a closed geometry.
If the skin thickness were over- or underestimated, we would obtain a density increase or
decrease towards the constriction, respectively.

\section{Pressure calculation}\label{sec:appB}

The pressure is calculated from the diagonal elements of the stress tensor which is calculated
for each individual particle $i$ as \cite{frenkel2001understanding,LAMMPS}
\begin{equation}\label{eq:StressTensor}
  S_{i a b}=-m_i v_{i a} v_{i b} -
\frac{1}{2}\sum_{\substack{j: p_j \in V_i\\j\neq i}} \left( r_{ia} F_{ijb} - r_{ja} F_{ijb}\right)
\end{equation}
with $a,b \in\{x,y,z\}$ the Cartesian components. The first term is
the ideal gas contribution and is biased by the collective flow speed.
Since only the thermal motion should contribute to $S_{i a b}$, the flow velocity
must be subtracted from $\vec{v}_i$, see section \ref{sec:appC} below for
the calculation of the flow velocity.
The second term is the virial contribution from the \gls{lj}-interaction.
The summation is over all particles $j$ within $r_c$ from
particle $i$, where $r_c$ is the cut-off radius of the LJ potential. This defines
the cut-off volume $V_i$ of particle $i$. 
$r_{ia}$ is component $a\in\{x,y,z\}$ of the coordinate of particle $i$ and
$F_{ijb}$ the component $b$ of the force of the pairwise interaction between particle
$i$ and $j$. We calculate the pressure $p(x)$ at position $x$ in the nozzle by
averaging the diagonal elements of the stress tensor $S_{i a b}$ over all particles $i$
within the bin volume $V_\mathrm{bin}(x)$,
\begin{equation}\label{eq:PressureFromStressTens}
  p(x)=
  -\left<\frac{\rho(x)}{3}\left(S_{i x x}+S_{i y y}+S_{i z z}\right)\right>_{t,V_\mathrm{bin}(x)}
\end{equation}
with $\left< \right>_{V_\mathrm{bin}(x)}$ denoting the average over $V_\mathrm{bin}(x)$.
We also average over the three diagonal elements because we
assume an isotropic stress tensor. Remembering that the temperature is not isotropic in the nozzle,
the assumption of an isotropic stress tensor may not be valid.
Inserting the stress tensor (\ref{eq:StressTensor}) into the expression
(\ref{eq:PressureFromStressTens}) for the local pressure, we obtain
\begin{align}
  p(x)&=
  \rho(x) k_\mathrm{B} T(x) +
  \frac{1}{3}\Bigg<\sum_{\substack{j:p_j \in \left(V_i \cap V_\mathrm{bin}(x)\right)\\j\neq i}}
  \!\!\!\!\!\!\!\!\!\!\textbf{r}_{i} \textbf{F}_{ij} \Bigg>_{\!\! t,V_\mathrm{bin}(x)}\nonumber\\
&+
  \frac{1}{6}\Bigg<\sum_{\substack{j:p_j \in \left(V_i \setminus V_\mathrm{bin}(x)\right)\\j\neq i}}
  \!\!\!\!\!\!\!\!\!\!\textbf{r}_{j} \textbf{F}_{ji} \Bigg>_{\!\! t,V_\mathrm{bin}(x)}
\label{eq:PressureFromStressTens5}
\end{align}
where in the calculation of the local virial we have to distinguish between
neighbor particles $p_j$ which are also in the same binning volume $V_\mathrm{bin}(x)$
as particle $p_i$ (giving rise to the first virial expression with the common prefactor
${1\over 3}$) and those which which are not (the second virial expression with
the prefactor ${1\over 6}$). For the first virial expression we could use
$\textbf{F}_{ij}=-\textbf{F}_{ji}$ and swap the summation index $i$ and $j$
leading to a factor 2. For the particles $p_j$ which are not in volume $V_\mathrm{bin}(x)$
this cannot be done, and each force $\textbf{F}_{ij}$ contributes just once.

\section{Calculation of Velocity}\label{sec:appC}

\begin{figure}[htbp]
	\begin{center}
		\includegraphics[width=1\columnwidth]{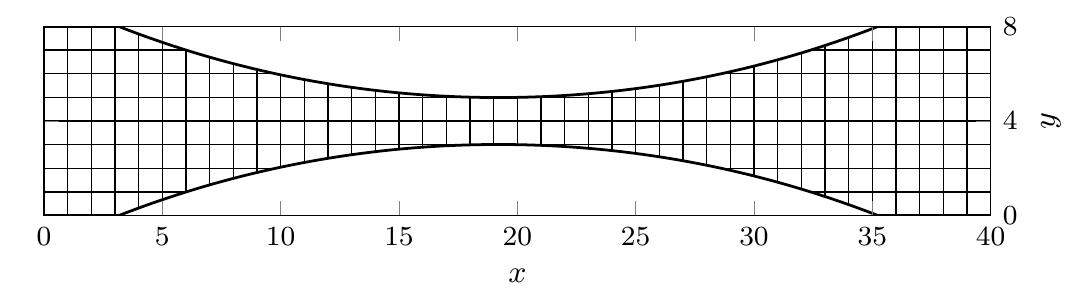}
	\end{center}
	\caption{Bin volumes $V_\mathrm{bin}(x,y)$ with side length $\sigma$
	in $x$- and $y$-direction.}
\label{VelocitySlices2}
\end{figure}

The velocity field $\textbf{v}(x,y)$
in the nozzle depends on both the $x$ and $y$-coordinate. The velocity is not only a
key quantity for Laval nozzles, but also required for obtaining the temperature $T$,
because $\textbf{v}(x,y)$ needs to be subtracted
from the particle velocities for the calculation of $T$, see next section. 
Fig.~\ref{VelocitySlices2} illustrates the bin volumes $V_\mathrm{bin}(x,y)$ for the
calculation of $\textbf{v}(x,y)$, as opposed to the bin slices in Fig.~\ref{DensitySlices2}.
The time averaged
flow velocity $\textbf{v}$ in a bin volume $V_\mathrm{bin}(x,y)$ can be calculated as
\begin{equation}\label{eq:VelocityAverage}
	v_a(x,y)=\left< {1\over N(x,y)} \sum_{i:p_i \in V_\mathrm{bin}(x,y)} v_{a i} \right>_{\!\! t}
\end{equation}
with $a\in\{x,y,z\}$, $v_{a i}$ is the velocity component $a$ of particle $p_i$,
and $N(x,y)$ the number of particles in $V_\mathrm{bin}(x,y)$ at a given time.
The magnitude of the flow velocity is
\begin{equation}\label{eq:ScalarVelocity}
	v(x)=\sqrt{\big<v_x(x,y)\big>^2_y+\big<v_y(x,y)\big>^2_y} 
\end{equation}
On average there is no flow in $z$-direction, $v_z(x,y)=0$.

\section{Temperature calculation}\label{sec:appD}

In order to investigate how the gas cools upon expanding supersonically through the nozzle,
we need to calculate the position-dependent temperature $T(x)$.
The microscopic definition of the temperature is the kinetic energy of the {\it random} part
of the particle velocity, hence we need to subtract the flow velocity $\textbf{v}(x,y)$
discussed in the previous section:
\begin{equation}
  k_\mathrm{B} T(x,y)=m \left< {1\over 3N(x,y)-3} \sum_{i:p_i \in V_\mathrm{bin}(x,y)}\!\!\!\!\!\!\!\!
  \left(\textbf{v}_{i} - \textbf{v}(x,y)\right)^2\right>_{\!\! t}
  \label{eq:TempCalc1}
\end{equation}
We are interested only in the $x$-dependence of the temperature and therefore we average
over $y$
\begin{equation}
  T(x) = \left<T(x,y)\right>_y
  \label{eq:TempCalc2}
\end{equation}
Note that subtracting the flow velocity removes three translational degrees of freedom, which we account
for by subtracting 3 from the number of degrees of freedom of the $N(x,y)$ particles in
binning volume $V_\mathrm{bin}(x,y)$.

In Eq.~(\ref{eq:TempCalc1}) we average over the contribution of the three velocity components,
which is fine in an isotropic system. In order to test whether the temperature is isoptropic
or not (and indeed we find it is not), we calculate the direction-dependent kinetic
temperature
\begin{equation}
  k_\mathrm{B} T_a(x,y)=m \left< {1\over N(x,y)-1} \sum_{i:p_i \in V_\mathrm{bin}(x,y)}\!\!\!\!\!\!\!\!
  \left(v_{ia} - v_a(x,y)\right)^2\right>_{\!\! t}
  \label{eq:TempCalc3}
\end{equation}
with $a\in\{x,y,z\}$. Again, we are interested only in how $T_a$ varies with position $x$
along the nozzle, hence we average over $y$, $T_a(x) = \left<T_a(x,y)\right>_y$.

\end{appendix}

\normalem
\bibliography{citations,hydro}
\end{document}